\documentclass[prd,aps,amsfonts,nofootinbib,notitlepage,longbibliography,10pt,superscriptaddress,twocolumn]{revtex4-1}

\usepackage{graphicx}
\usepackage{tabularx}
\usepackage{braket}
\usepackage{amssymb,mathtools}
\usepackage{amsmath}
\usepackage{bm}
\usepackage[driverfallback=dvipdfm]{hyperref}
\hypersetup{colorlinks=true,breaklinks,urlcolor=blue,linkcolor=blue,citecolor=blue}

\usepackage{tikz}
\usetikzlibrary{decorations.pathreplacing}
\usetikzlibrary{arrows.meta}
\usetikzlibrary{patterns}

\usepackage{pdfpages}
\usepackage{tikz}
\makeatletter
\AtBeginDocument{\let\LS@rot\@undefined}
\makeatother

\newcommand{\revise}[1]{#1}

\newcommand{\comment}[1]{}
\usepackage{xcolor}
\definecolor{ginger}{rgb}{0.69, 0.4, 0.0}

\newcommand{\lr}[1]{\left( #1\right)}
\newcommand{\mlr}[1]{\left[ #1\right]}
\newcommand{\glr}[1]{\left\{ #1\right\}}
\newcommand{\alr}[1]{\left\langle #1\right\rangle}
\newcommand{\norm}[1]{\left\lVert#1\right\rVert}
\newcommand{\abs}[1]{\left\lvert#1\right\rvert}

\newcommand{\ii}{\mathrm{i}}
\newcommand{\ee}{\mathrm{e}}
\newcommand{\dd}{\mathrm{d}}

\newcommand{\trace}{\mathrm{Tr}}

\newcommand{\where}{\quad {\rm where}\quad}
\newcommand{\order}{\mathrm{O}}

\newcommand{\floor}[1]{\left\lfloor#1\right\rfloor}

\newcommand{\poly}[1]{\mathrm{poly}\lr{#1} }

\newcommand{\OO}{\mathcal{O}}

\newcommand{\kpsi}{\ket{\psi}}

\newcommand{\bz}{\mathbf{z}}
\newcommand{\bs}{\mathbf{s}}
\newcommand{\prand}{p}
\newcommand{\mytag}[1]{\tag*{\llap{$\lr{#1}$}}}

\renewcommand{\thesubsection}{\thesection.\arabic{subsection}}

\makeatletter

\renewcommand\onecolumngrid{
\do@columngrid{one}{\@ne}%
\def\set@footnotewidth{\onecolumngrid}
\def\footnoterule{\kern-6pt\hrule width 1.5in\kern6pt}%
}

\renewcommand\twocolumngrid{
        \def\footnoterule{
        \dimen@\skip\footins\divide\dimen@\thr@@
        \kern-\dimen@\hrule width.5in\kern\dimen@}
        \do@columngrid{mlt}{\tw@}
}%

\renewcommand{\p@subsection}{}
\renewcommand{\p@subsubsection}{}
\makeatother

\usepackage{amsthm}
\newtheorem{thm}{Theorem}
\numberwithin{thm}{section}
\newtheorem{cor}[thm]{Corollary}

\newtheorem{prop}[thm]{Proposition}
\newtheorem{eg}[thm]{Example}

\newtheorem{defn}[thm]{Definition}

\newtheorem{rmk}[thm]{Remark}

\begin{document} 

\title{Robust symmetry breaking in gapless quantum magnets}
\author{Chao Yin}\email{chaoyin@stanford.edu}
\affiliation{Department of Physics, Stanford University, Stanford, California 94305, USA}

\author{Andrew Lucas}\email{andrew.j.lucas@colorado.edu}
\affiliation{Department of Physics and Center for Theory of Quantum Matter, University of Colorado, Boulder, Colorado 80309, USA}

\date{\today}

\begin{abstract}
 We prove the existence of spontaneous symmetry breaking in suitably low-energy eigenstates of certain gapless and frustrated many-body quantum systems, namely symmetric quantum perturbations to classical models which exhibit spontaneous symmetry breaking of a finite group at some positive temperature.  Additionally, the classical model need not be local in space, as long as it satisfies a quantum analogue of the Peierls condition. As an example of our technique, we establish robust ferromagnetism in random-bond Ising models in $d= 2$ dimensions with sufficiently biased random couplings, with weak transverse field.   Our mathematical technique is based on establishing quantum bottlenecks, similar to a ``many-body WKB" method for evaluating tunneling rates.  Using these same methods, we provide new proofs of metastability and the slow decay of the false vacuum, applicable to gapless metastable states.  Our work represents a first step towards a rigorous classification of stable gapless quantum phases.
\end{abstract}

\maketitle

\emph{Introduction.}---
 The paradigm of spontaneous symmetry breaking (SSB) is fundamental across many-body physics.   As a simple illustration of SSB, consider a model of $N$ interacting Ising spins $z_i\in \lbrace \pm 1 \rbrace$, with classical Hamiltonian \begin{equation}
     H_0 = \sum_{i,j=1}^N J_{ij}z_i z_j.
 \end{equation}
 $H_0$ is invariant under the transformation $z_i \rightarrow -z_i$ for each $i$, which is called a $\mathbb{Z}_2$ symmetry of the problem.   If $\mathbf{z}_0$ is a configuration which minimizes $H_0$, so is $-\mathbf{z}_0$.  These two configurations are mapped on to each other by the action of the global symmetry.  As such, the two states $\pm \mathbf{z}_0$ spontaneously break the $\mathbb{Z}_2$ symmetry of $H_0$.

 In many-body quantum systems we might also expect SSB, but as is well-known, defining it carefully is a bit more subtle \cite{andersontower, ginibre, frohlich, Koma_1994}.  For example, we may study a quantum Ising model where the Hamiltonian $H$ has non-commuting terms, albeit symmetric under the global $\mathbb{Z}_2$ (spin flip) symmetry described above.   In this case, we expect the quantum ground state to take a schematic ``Schr\"odinger cat" form \begin{equation}
     |\text{gs}\rangle \sim |\mathbf{z}_0\rangle + |-\mathbf{z}_0\rangle.
 \end{equation}
 Here, the intuition is that the state exhibits SSB because it is in a superposition of two macroscopically different wave functions, \revise{each component of which exhibits classical long-range order: $\abs{\langle \mathrm{gs}|Z_i Z_j |\mathrm{gs}\rangle}$ = 1.}   The macroscopic superposition is a mathematical requirement \revise{(eigenstates of $H$ transform in irreducible representations of the symmetry $X_{\mathrm{tot}}$ which flips all spins, and indeed $X_{\mathrm{tot}}|\mathrm{gs}\rangle = |\mathrm{gs}\rangle$), but any local correlation function $\langle \mathrm{gs}|A|\mathrm{gs}\rangle = \langle \mathbf{z}_0|A|\mathbf{z}_0\rangle + \langle -\mathbf{z}_0|A|-\mathbf{z}_0\rangle$ is equivalent to that in an incoherent mixed state of $|\mathbf{z}_0\rangle$ and $|-\mathbf{z}_0\rangle = X_{\mathrm{tot}}|\mathbf{z}_0\rangle$.}

 In this letter, we are interested in mathematically establishing the existence of SSB in low-energy eigenstates of many-body quantum systems.  We present a new proof  inspired by many-body eigenstate localization \cite{MBL_LDPC24} and quantum bottleneck theory \cite{Gamarnik:2024hyk,q_bottle_Vedika}.  Our methods, roughly speaking, establish the existence of SSB in quantum models which are perturbations of a commuting (classical) model where a discrete symmetry group is spontaneously broken at some positive temperature.    As long as the perturbations locally respect the symmetry, we will show how to leverage the classical Peierls argument to establish the stability of SSB in a low-energy many-body eigenstate.  We say these models obey a \emph{quantum Peierls condition} (QPC); we will discuss this more concretely later.

 What is most interesting about our new proof is that it does not rely on the many-body state being gapped, or the parent Hamiltonian being frustration-free. These are two standard assumptions underlying earlier proofs that a quantum phase of matter is stable \cite{yarotsky2006ground,topo_Hastings,bravyi2011short,michalakis2013stability,frustration_free_22,yin_ldpc,ldpc_roeck}.   We can also establish the robustness of SSB for certain non-local Hamiltonians, while previous proof techniques require the unperturbed Hamiltonian to be short-range at least in some bounded-degree graph \cite{michalakis2013stability,yin_ldpc,ldpc_roeck}. We stress that the very \emph{definition} of a gapped phase relies on an adiabatic path between ground states that does not close the gap \cite{QI_meet_QM}; an entirely new paradigm is needed to classify phases of gapless matter, let alone establish their stability. Our techniques may prove useful in this regard, as we can establish that SSB is robust to symmetric perturbations of gapless and frustrated models obeying the QPC. For example, we establish robust ferromagnetism in quantum random-bond Ising models (RBIM) in $d=2$ dimensions \cite{RBIM_exp99,RBIM98,RBIM_PRL98,infiniterandomness,RBIM_rev05,RBIM06,RBIM_RG10,RBIM25,RBIM26}, confirming a longstanding conjecture in quantum statistical mechanics.  (We expect the extension to $d>2$ is easy \revise{following e.g. \cite{lattice_book17}}, but do not present a proof.)

 Another helpful aspect of our formalism is that the thermodynamic limit plays only a modest role in our method. This allows us to, in principle, provide upper bounds on the system size needed before various signatures of SSB could be observed with finitely many degrees of freedom: e.g. lower bounds on the time scale where the symmetry can appear restored in an SSB state.  This perspective may be helpful when considering the modest system sizes accessible in modern quantum simulators.  

\emph{Formalism.}---
We now describe our formalism at a high level, focusing on a special case for pedagogical purposes; the case \revise{of general discrete symmetries} is presented in SM \cite{SM}. Consider $N$ interacting qubits labeled by $i\in\lbrace 1,\ldots , N\rbrace$; let $X_i, Y_i, Z_i$ denote the Pauli matrices acting on qubit $i$.  We are interested in Hamiltonians $H$ that are invariant under the global $\mathbb{Z}_2$ symmetry generated by $X_{\mathrm{tot}} = X_1\cdots X_N$:  $[H,X_{\mathrm{tot}}]=0$.   In particular, let us consider $H=H_0+V$, where \begin{equation}
    H_0 = -\sum_{i<j}J_{ij}Z_iZ_j, \;\;\; V = \epsilon \sum_i X_i. \label{eq:quantumising}
\end{equation}
\eqref{eq:quantumising} is the well-studied transverse-field Ising model.  The essential properties that we need are (\emph{i}) $H_0$ is classical: each term only depends on Pauli $Z$;  (\emph{ii}) each term in $H_0$ and $V$ commute with $X_{\mathrm{tot}}$ individually.   We will also assume that (\emph{iii}) the couplings $J_{ij}$ are chosen to obey a classical \emph{Peierls condition} (PC).

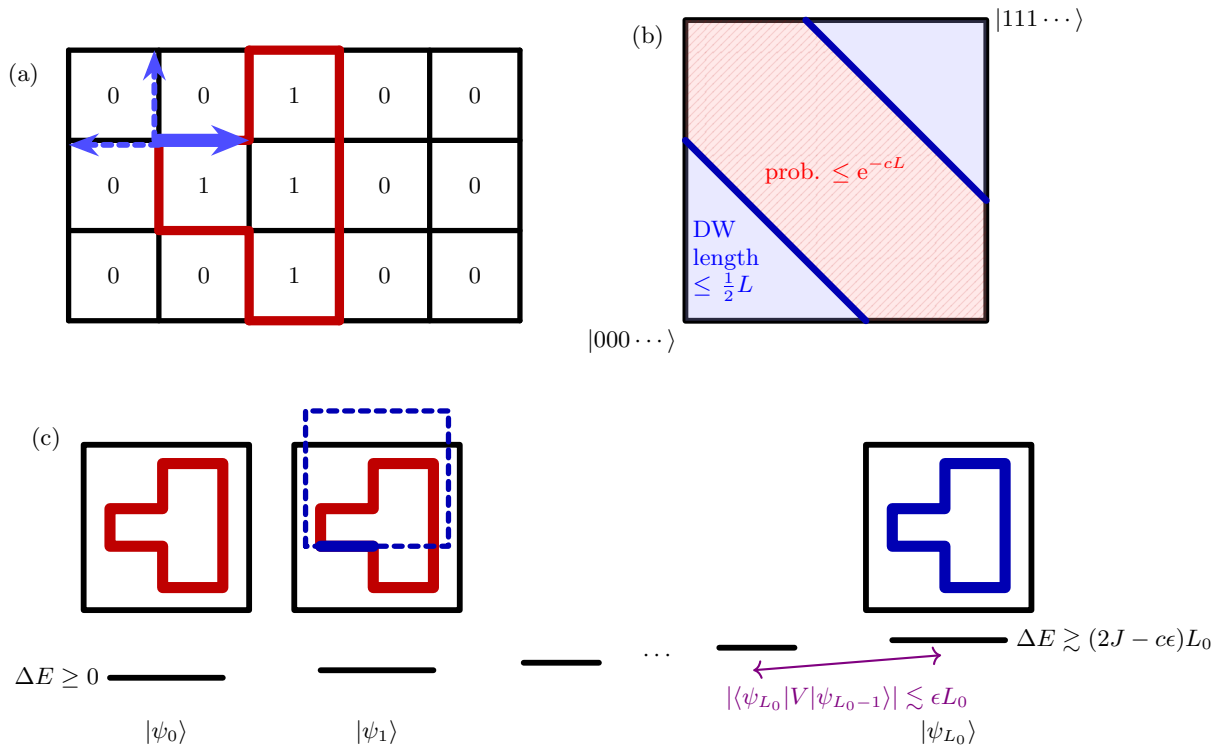
\begin{figure*}
    \centering

    \begin{tikzpicture}

\begin{scope}[scale=1.2, line cap=round, line join=round]

  \tikzset{
    grid/.style={black, line width=2.0pt},
    rededge/.style={red!75!black, line width=3.5pt},
    blueSolid/.style={blue!70!white, -{Stealth[length=5mm,width=4mm]}, line width=5pt},
    blueDash/.style={blue!70!white, dashed, -{Stealth[length=4mm,width=3mm]}, line width=2.2pt},
  }

\draw (-1.5, 2.5) node[above] {(a)};
  \foreach \x in {-1,0,1,2,3,4} { \draw[grid] (\x,0) -- (\x,3); }
  \foreach \y in {0,1,2,3}      { \draw[grid] (-1,\y) -- (4,\y); }


  \draw[rededge] (1,0) -- (2,0);   
  \draw[rededge] (2,0) -- (2,3);   
  \draw[rededge] (2,3) -- (1,3);   

  \draw[rededge] (0,1) -- (0,2);   
  \draw[rededge] (0,1) -- (1,1) -- (1,0);   
  \draw[rededge] (0,2) -- (1,2) -- (1,3);   

  \draw[blueSolid] (0,2) -- (1,2);     
  \draw[blueDash]  (0,1.95) -- (-1,1.95);    
  \draw[blueDash]  (-0.05,2) -- (-0.05,3);     

  \draw (-0.5, 0.5) node {0};
  \draw (0.5, 0.5) node {0};
  \draw (1.5, 0.5) node {1};
  \draw (2.5, 0.5) node {0};
  \draw (3.5, 0.5) node {0};

    \draw (-0.5, 1.5) node {0};
  \draw (0.5, 1.5) node {1};
  \draw (1.5, 1.5) node {1};
  \draw (2.5, 1.5) node {0};
  \draw (3.5, 1.5) node {0};

    \draw (-0.5, 2.5) node {0};
  \draw (0.5, 2.5) node {0};
  \draw (1.5, 2.5) node {1};
  \draw (2.5, 2.5) node {0};
  \draw (3.5, 2.5) node {0};

\end{scope}


\begin{scope}[line cap=round, line join=round, xshift=7cm]

\tikzset{
  box/.style={black, line width=2.2pt},
  bluereg/.style={fill=blue!35, fill opacity=0.25, draw=none},
  bottleneck/.style={fill=red!35, fill opacity=0.25, draw=none},
  hatch/.style={pattern=north east lines, pattern color=red!70!black},
  blueedge/.style={blue!70!black, line width=2.6pt},
  arrow/.style={-{Stealth[length=4mm,width=3mm]}, line width=1.8pt},
  darrow/.style={dashed, -{Stealth[length=4mm,width=3mm]}, line width=1.6pt},
}

\draw (-0.5, 3.5) node[above] {(b)};

\draw[box] (0,0) rectangle (4,4);

\coordinate (A) at (1.6,4);  
\coordinate (B) at (4,1.6);  
\coordinate (C) at (0,2.4);  
\coordinate (D) at (2.4,0);  

\fill[bluereg] (A) -- (4,4) -- (B) -- cycle;

\fill[bluereg] (0,0) -- (D) -- (C) -- cycle;

\fill[bottleneck] (A) -- (B) -- (4,0) -- (D) -- (C) -- (0,4) -- cycle;

\fill[hatch, opacity=0.35] (A) -- (B) -- (4,0) -- (D) -- (C) -- (0,4) -- cycle;

\draw[blueedge] (A) -- (B);
\draw[blueedge] (C) -- (D);

\node[anchor=north east] at (0,0) {$|000\cdots\rangle$};

\draw (4,4) 
  node[anchor=west] {$|111\cdots\rangle$};

\draw (0.6,0.8) node[text width=1cm] {\color{blue}DW length $\le \frac{1}{2}L$};

\draw (2,2) node {\color{red} prob. $\le \mathrm{e}^{-cL}$};

\end{scope}


\begin{scope}[line cap=round, line join=round, xshift=-1cm, yshift=-3.85cm]

\tikzset{
  box/.style={black, line width=2.2pt},
  rpath/.style={red!75!black, line width=4.2pt},
  bpath/.style={blue!70!black, line width=4.2pt},
  bdash/.style={blue!70!black, dashed, line width=2.0pt},
  bar/.style={black, line width=2.2pt},
  bluearrow/.style={blue!70!black, -{Stealth[length=4mm,width=3mm]}, line width=4.2pt},
}

\draw (-0.5, 2) node[above] {(c)};

\def\S{2.2}      
\def\gap{2.8}    

\def\xL{0.35}
\def\xM{1.05}
\def\xR{1.85}
\def\yB{0.30}
\def\yM{0.85}
\def\yMM{1.35}
\def\yT{1.95}

\newcommand{\drawRedMaze}{%
  \draw[rpath]
    (\xL,\yM) -- (\xL,\yMM) -- (\xM,\yMM) -- (\xM,\yT) -- (\xR,\yT) -- (\xR,\yB) -- (\xM,\yB) -- (\xM,\yM) -- cycle;
}

\newcommand{\drawBlueMaze}{%
  \draw[bpath]
    (\xL,\yM) -- (\xL,\yMM) -- (\xM,\yMM) -- (\xM,\yT) -- (\xR,\yT) -- (\xR,\yB) -- (\xM,\yB) -- (\xM,\yM) -- cycle;
}

\begin{scope}[shift={(0,0)}]
  \draw[box] (0,0) rectangle (\S,\S);
  \drawRedMaze;
  \draw[bar] (0.35,-0.9) node[left] {$ \Delta E \ge 0$} -- (\S-0.35,-0.9);
  \node[anchor=north] at (\S/2,-1.35) {$|\psi_0\rangle$};
\end{scope}

\begin{scope}[shift={(\gap,0)}]
  \draw[box] (0,0) rectangle (\S,\S);
  \drawRedMaze;

  \draw[bdash] (\xL-0.2,\yM) rectangle (2.05,2.65);

  \draw[bpath] (\xM,\yM) -- (\xL,\yM);

  \draw[bar] (0.35,-0.8) -- (\S-0.35,-0.8);
  \node[anchor=north] at (\S/2,-1.35) {$|\psi_1\rangle$};
\end{scope}

\draw[bar] (2*\gap+0.25,-0.7) -- (2*\gap+1.25,-0.7);
\node at (2*\gap+2.05,-0.6) {$\cdots$};
\draw[bar] (2*\gap+2.85,-0.5) -- (2*\gap+3.85,-0.5);

\begin{scope}[shift={(2*\gap+4.8,0)}]
  \draw[box] (0,0) rectangle (\S,\S);
  \drawBlueMaze;

  \draw[bar] (0.35,-0.40) -- (\S-0.35,-0.40) node[right] {$\Delta E \gtrsim  (2J-c\epsilon)L_0$};
  \node[anchor=north] at (\S/2,-1.35) {$|\psi_{L_0}\rangle$};
\end{scope}

\draw[<->, violet, thick] (2*\gap+5.8, -0.6) -- (2*\gap+3.3, -0.8);
\draw (2*\gap+4.55,-0.9) node[below] {\color{violet} $|\langle \psi_{L_0}|V|\psi_{L_0-1}\rangle | \lesssim \epsilon {L_0}$};

\end{scope}

\end{tikzpicture}

    \caption{(a) Domain walls specify excitations about a ground state; the number of distinct domain walls of length $\ell$ passing through a given vertex on the square lattice is $<3^\ell$, as illustrated by the blue step (there were 3 possible places for the domain wall to grow).  (b) The classical Peierls condition implies a bottleneck in the configuration space -- long domain walls are exceedingly hard to find at low temperature in the classical Gibbs ensemble.  (c) The quantum Peierls condition implies that a low-energy many-body wave function is very unlikely to be found in a state where all of the bonds along a large closed loop are flipped.  In red we illustrate a fixed loop $\gamma$, while blue edges represent examples of wave functions that exist in each sector.  The dashed blue line denotes one possible way that such an excitation can exist (which we do \emph{not} keep track of in \eqref{eq:cleverbasis}). $\Delta E$ is the expectation of $H-E_0$ in the sector.}
    \label{fig:maintext}
\end{figure*}

The classical PC is easy to introduce in the context of the classical 2d Ising model, where $J_{ij}=J$ are non-zero (and positive) only when $i$ and $j$ are nearest neighbors on a 2d square lattice.  Here, the SSB ground states of $H_0$ have either all $Z_i=1$ or $Z_i=-1$.    As illustrated in Figure \ref{fig:maintext}(a), we can track excitations above the ground state of $H_0$ by keeping track of nearest neighbors where $Z_iZ_j=-1$.  These ``domain walls" between regions of $\pm 1$ spin form closed loops.   Suppose a closed loop of length $\ell$ passes through some vertex $i$:  there are $<3^\ell$ loops possible (when drawing a loop, there are $\le 3$ options for where to place the next edge), while each loop has an energy cost of $2J\ell$. In the thermal ensemble where each spin configuration $\mathbf{z}$ is observed with probability $\propto \mathrm{e}^{-\beta H_0(\mathbf{z})}$, the probability of seeing one specific domain wall is $\sim \mathrm{e}^{-2\beta J \ell}$ at large $\ell$.  The probability of seeing \emph{any} domain wall of length $\ell$ pass through vertex $i$ is bounded by $(3\mathrm{e}^{-2\beta J})^\ell $, and this becomes exponentially small with $\ell$ once $2\beta J>\log 3$. Since there are only $N$ choices of vertex $i$,  we readily conclude that the probability of seeing a system-size-spanning domain wall is exponentially small in the system length $L\sim \sqrt{N}$ which is negligible at low temperature.   As depicted in Figure \ref{fig:maintext}(b), we can therefore deduce that typical states drawn from a low-temperature thermal ensemble are ferromagnetic, with long-range order and $\mathbb{Z}_2$ SSB.   The small density of flipped spins form small disconnected clusters, and the probability of seeing a large cluster exponentially vanishes. The states with a large cluster \revise{(and therefore a large domain wall)} constitute the \emph{bottleneck} \revise{space $\Phi$} responsible for the exponential slowdown of equilibration in a thermal bath \cite{levin_markovchains}. \revise{More precisely, $\Phi$ includes all states $\bz$ such that $P_\gamma \ket{\bz}=\ket{\bz}$ for some domain wall $\gamma$ of length $\gtrsim L$, where \begin{equation}
    P_\gamma = \prod_{i\sim j\text{ along }\gamma } \frac{1-Z_iZ_j}{2}.
\end{equation}
}This thermally stable SSB relies on the PC: large domain walls are energetically more costly than they are entropically favorable.

Now let us return to the quantum many-body problem.  The 2d transverse-field Ising model $H=H_0+V$ at $\epsilon\ll J$ is known to have two nearly degenerate low-energy states exhibiting SSB separated by a finite spectral gap from the rest of the spectrum.  Nevertheless, it will be instructive to re-derive the existence of SSB using the classical PC. Earlier literature \cite{ginibre,spacetime_CE85,spacetime_CE96,spacetime_CE_96} has established how to use the classical PC to establish SSB in the thermal partition function; in contrast, we will be directly studying the low-lying eigenstates of $H$ following the modern literature on stable quantum phases, where we only need PC for large-scale excitations.  We will summarize our perspective here, leaving technical details to SM.   

Our goal is to show there are two nearly-degenerate ground states $|\Psi_{1,2}\rangle$ whose eigenvalues are quite close: \begin{equation}
    H|\Psi_1\rangle = E_0 |\Psi_1\rangle, \quad H|\Psi_2\rangle = (E_0 +\delta E_0) |\Psi_2\rangle
\end{equation}
with $\delta E_0 \lesssim \mathrm{e}^{-L}$ (this is weaker than the actual $\lesssim \mathrm{e}^{-L^2}$ energy difference \cite{longrange_stab23,yin_ldpc,ldpc_roeck,LRB_volumetail}).  This means that $|\pm\rangle=\ket{\Psi_1}\pm \ket{\Psi_2}$ alone are ``almost eigenstates": \begin{equation}
    \lVert (H-E_0)|\pm\rangle \rVert \lesssim \mathrm{e}^{-L}.\label{eq:mainalmost}
\end{equation}  We can anticipate that the almost eigenstates $|\pm\rangle$ are short-range entangled states made out of quantum superpositions of small clusters of flipped spins -- much like how the typical low-temperature states of the classical model correspond to a sea of $\pm$ spins with small clusters of $\mp$ spins inside.  Indeed, we make this precise, by calculating the fraction of the wave function $|\pm\rangle$ that contains a specific (large) domain wall $\gamma$ of length $\ell$: we aim to bound, for large enough $\ell$, \begin{equation}
    \lVert P_\gamma |\pm\rangle \rVert \lesssim (c^\prime \epsilon/J)^{c\ell } \label{eq:main_bottleneck}
\end{equation}
for O(1) constants $c$ and $c^\prime$, when $\epsilon$ is sufficiently small.

The intuition for this result is simple: each individual $X_i$ in $V$ can flip at most 4 bonds, so we need to go to order $\frac{\ell}{4}$ in perturbation theory to flip the entire domain wall, suggesting a suppression factor of $\epsilon^{\ell/4}$.  However, we wish to avoid this naive perturbative argument.  The perturbation theory involves energy denominators $\epsilon^n/\mathrm{\Delta} E^{n-1}$, and we do not know the many-body spectrum to compute $\mathrm{\Delta}E$.  More importantly, the Peierls argument does \emph{not} care about relative energy-level spacings, only on the ``final" energy barrier of a large cluster.

Therefore, we will use a different technique inspired by many-body eigenstate localization \cite{MBL_LDPC24}, or alternatively a many-body WKB/tunneling picture.  The key point is as follows: if $P_\gamma |\psi\rangle = |\psi\rangle$, then construct the state $|\psi^\prime\rangle = X_\gamma |\psi\rangle$, where $X_{\gamma}$ is the product of Pauli $X$ operators inside the domain wall traced by $\gamma$.   One can readily calculate that \begin{equation}
    E_0 \le \langle \psi^\prime|H|\psi^\prime \rangle = \langle \psi|H|\psi\rangle - (2J-c\epsilon )\ell \label{eq:mainQPC}
\end{equation}
for some constant $c$.
Here we have used $E_0$ is the ground energy, the PC for the energy difference $2J\ell$, and that $V$ is local and symmetric so the truncated symmetry operator $X_\gamma$ changes energy of $V$ by at most $\sim \epsilon \ell$ near $\gamma$ (For the special case of transverse-field, $\alr{V}$ is not changed at all).
From \eqref{eq:mainQPC}, $|\psi\rangle$ is a highly excited state when $\epsilon \ll J$.  A similar argument shows that any state with more than half of the checks ($Z_iZ_j$) along $\gamma$ violated is comparably excited.  With this in mind, we \revise{expand} in a clever basis \cite{MBL_LDPC24} \begin{equation}
    |+\rangle = \sum_{n=0}^\ell  A_n |\psi_n\rangle \label{eq:cleverbasis}
\end{equation}
where each $|\psi_n\rangle$ is a normalized state which contains the states that have exactly $n$ flipped checks along $\gamma$. \revise{Plugging \eqref{eq:cleverbasis} into the many-body Schr\"odinger equation $H|+\rangle \approx  E_0|+\rangle$ [recall \eqref{eq:mainalmost}],} the ``hopping'' $V$ only couples nearby $n$ due to locality, so\begin{align}\label{eq:iterate}
    \langle \psi_n| (H_0+V-E_0) |\psi_n\rangle A_n \approx  -\sum_{n^\prime=n-4}^{n+4} \langle \psi_n| V|\psi_{n^\prime}\rangle A_{n^\prime}.
\end{align}
Generalizing the above arguments we can show that for all $n>n_*=\frac{2}{3}\ell$,  $\langle \psi_n| (H_0+V-E_0) |\psi_n\rangle \gtrsim (J-c\epsilon)\ell $ while $\langle \psi_n| V|\psi_{n^\prime}\rangle \lesssim \epsilon \ell$. This equation therefore describes a \emph{many-body} wave function trapped inside of a bottleneck \revise{as in \cite{MBL_LDPC24}: \eqref{eq:iterate} at $n=\ell$ implies $A_\ell\lesssim (\epsilon/J) \widetilde{A}_{\ell-1}$ with $\widetilde{A}_{\ell-1}=\sqrt{\sum_{m=1}^4 A_{\ell-m}^2}$ being the amplitude of a group of $4$ neighboring $n$s. We then investigate \eqref{eq:iterate} at $n=\ell-1,\cdots$ to obtain $\widetilde{A}_{\ell-1}\lesssim (\epsilon/J) \widetilde{A}_{\ell-2}$ bounded by the amplitude of the next group, and so on. Iterating this yields} for $n>n_*$ \begin{equation}\label{eq:main_An<}
    A_n \lesssim \left(\frac{\epsilon}{J}\right)^{(n-n_*)/4}
\end{equation}
which immediately implies \eqref{eq:main_bottleneck}.

Now, the probability that $|+\rangle$ has \emph{any} weight near the top of the classical bottleneck \revise{$\Phi$} is bounded by \begin{equation}\label{eq:maintext_bottleneck}
    \sum_{\gamma \text{ of length }  L_0} \left\lVert  P_\gamma |\psi\rangle \right\rVert \le L^2 3^{L_0}\left(\frac{c'\epsilon}{J}\right)^{cL_0}.
\end{equation}
Here we can take $L_0\sim L$ to  scale with the system size, and we easily see that for large $L$ and small $\epsilon$, the many-body wave function $\ket{+}$ is trapped inside the bottleneck implied by the Peierls condition. Like the low-temperature case, the energy suppression \eqref{eq:main_An<} for a given large domain wall dominates the ``entropy'' of the number of domain walls. As a consequence, we can establish \eqref{eq:mainalmost}, which itself implies that expectation values of observables, $\langle \pm| A(t)|\pm\rangle $, appear almost frozen for a time scale $t_{\mathrm{SSB}} \gtrsim \mathrm{e}^L$. With a more careful investigation of the bottleneck states, we are able to further prove in SM that $\ket{+}$ ($\ket{-}$) has macroscopic positive (negative) polarization $\sum_i Z_i$, justifying SSB.

 The essential lynchpin in the argument above is the energy estimate \eqref{eq:mainQPC}, a special case of what we call the \emph{quantum Peierls condition} (QPC):  the existence of bottlenecks in the many-body Hilbert space where we can guarantee that every wave function in the bottleneck is highly excited. Moreover, QPC for the perturbed $H$ comes from a PC for the energy landscape of the classical $H_0$. In SM, we show that a QPC is sufficient to establish the existence of almost eigenstates with ultra-slow dynamics.   The fact that $H_0$ was gapped, frustration-free or short-range does not enter into the proof at all, in contrast to previous methods \cite{yarotsky2006ground,topo_Hastings,bravyi2011short,michalakis2013stability,frustration_free_22,yin_ldpc,ldpc_roeck} based on perturbation theories.   Indeed, we will now prove the existence of SSB in models which are disordered, gapless and frustrated.

\emph{Applications.}---
 To showcase the broad implications of our new proof, let us establish the stability of a number of variants of quantum Ising models.   First, we consider the random-bond quantum Ising model, where the couplings $J_{ij}$ in \eqref{eq:quantumising} are again only non-zero on the edges of a 2d square lattice, but now are random variables.  As long as this distribution is bounded and $J_{ij}$ is biased towards being ferromagnetic, we verify that the QPC for large domain walls still holds with overwhelming probability as we increase the size of the lattice.  Therefore, an identical proof to the one sketched above establishes that the ground states of the quantum random-bond Ising model \revise{at sufficiently small perturbation strength $\epsilon$} are ferromagnetic (namely they have SSB); this rigorously verifies a longstanding expectation \cite{RBIM_exp99,RBIM98,RBIM_PRL98,infiniterandomness,RBIM_rev05,RBIM06,RBIM_RG10,RBIM25,RBIM26}. \revise{Similar to other perturbation-based proofs \cite{yarotsky2006ground,topo_Hastings,bravyi2011short,michalakis2013stability,frustration_free_22,yin_ldpc,ldpc_roeck}, the actual phase may extend well beyond the range of $\epsilon$ for which our proof works. }  
 
Our proof of SSB only holds sufficiently far away from the critical field strength where the ferromagnetism is lost.  Close to this critical field strength, we may think of the Ising model as near the quantum critical point of a 2+1-dimensional $\phi^4$ theory; random bonds correspond to random fluctuations in $\phi^2$, which is a Harris-relevant operator.  It is therefore expected that the critical point is drastically modified by disorder \cite{harris}. 

As a second example, we now turn to an Ising model on an $L\times L$ square lattice where the flavor $\mathbb{Z}_2$ symmetry is mixed with translation: \begin{equation} \label{eq:evenodd}
    H = -J\sum_{i\sim j}Z_iZ_j + h\left[\sum_{i \in \text{even}} Z_i - \sum_{i \in \text{odd}} Z_i\right] + V
\end{equation}
where even/odd refer to the two halves of the bipartite square lattice.  In general, $h\ne 0$ corresponds to a symmetry breaking perturbation which destroys ferromagnetism.  However, we can actually prove that ferromagnetism is stable in this theory so long as $V$ is symmetric under the mixed flavor/translation symmetry. From a field theory perspective, evidently the $h$ perturbation in \eqref{eq:evenodd} is not a relevant operator, which is non-trivial as the sign change between even/odd sites is naively simply a finite wave number modification to an otherwise relevant perturbation.  It is expected that ferromagnetism is \emph{not} stable if e.g. the perturbation is symmetric under $\mathbb{Z}_2$ but not translation:  for example, adding the perturbation $\epsilon \sum_{i\in \text{even}}X_i$ is expected to cause the $\langle Z_i\rangle > 0$ sector to be preferred.   

Although the discussion above highlights the role of symmetry in establishing a QPC, we stress that the QPC does not rely on symmetry. As a simple illustration thereof, consider modifying a 2d quantum Ising model by adding a small longitudinal field: \begin{equation} \label{eq:maintext_FV_H}
    H = H_0 +V - h\sum_i Z_i
\end{equation} This single perturbation does not spoil the QPC as long as perturbation $V$ remains locally symmetric, and $h \ll 1/L$. After explicitly breaking the symmetry, we are no longer guaranteed that exact eigenstates of $H$ transform in irreducible representations of $\mathbb{Z}_2$.  One may then ask whether the eigenstates cluster in mostly positive or mostly negative magnetization states.  It is widely expected that this will indeed happen for the ground state, even if $h\sim \mathrm{e}^{-L}$.   Using the same methods sketched above, we are indeed able to establish this fact in SM.  

  In classical statistical physics, the notion of ``clustering" persists to a finite temperature or energy density.  If this persists at finite energy density in quantum systems, it would arguably violate the eigenstate thermalization hypothesis \cite{srednicki,Deutsch:2018ulr}; some evidence for this in small systems is presented in \cite{srednicki15}.  

At larger values of $h$, \eqref{eq:maintext_FV_H} will exhibit a false vacuum \cite{coleman,FVD_rev22,falseVac_Ising,falseVac_spinchain,our_preth,metastable_localgap,falseVac_confine,Bubble2021,Pomponio22,Milsted22,lagnese2023detecting, FVDTakacs, FVDphi4, Maki23, Batini,ferrari,darbha2024false,vodeb2024stirring}, which may be gapless in e.g. the quantum RBIM in $d\ge 2$.   Once $h\ne 0$, the global $\mathbb{Z}_2$ symmetry is explicitly broken, so the vacuum should be unique and have $\langle Z_i\rangle >0$ for a typical site.  The typical picture in physics is that if we start in an initial state with most spins in $\langle Z_i\rangle <0 $, this is a finite energy-density state and eventually should thermalize; however, we must nucleate a critical bubble of radius $R_{\mathrm{c}}\sim J/h$ in order to destabilize the false vacuum.  Hence the decay of the false vacuum should be non-perturbatively slow.  When every $J_{ij}>\Delta$, the above Hamiltonian is gapped and a  rigorous analysis of the false vacuum decay problem was developed in \cite{our_preth,metastable_localgap}, following earlier work on prethermalization in e.g. the Hubbard model \cite{abanin2017rigorous}: in particular, the false vacuum $|\psi\rangle$ was identified with a specific many-body state, and the dynamics of local correlators such as $\langle \psi| Z_i(t)|\psi\rangle$ in the false vacuum was established.  

To apply our technique to the false vacuum problem in e.g. the 2d qRBIM, the key observation is that the QPC continues to hold for domain walls whose length $\ell<L_0\sim R_{\mathrm{c}}$.   If we restricted the dynamics to an $L^\prime \times L^\prime$ lattice where $L^\prime \lesssim \mathrm{e}^{R_{\mathrm{c}}}$, then the analysis leading to \eqref{eq:maintext_bottleneck} is unchanged: we could construct a false vacuum wave function with most $\langle Z_i\rangle < 0$ which is an almost eigenstate of $H$ in \eqref{eq:maintext_FV_H}.  $L^\prime$ is a non-perturbatively large, but still finite, number.   When the true system size $L \gg L^\prime$, the false vacuum is no longer a \emph{global} almost eigenstate, but it remains the case \emph{local} observables such as $\langle \psi| Z_i(t)|\psi\rangle$ are only sensitive to a ball of radius $vt$ surrounding site $i$:  here $v$ is the Lieb-Robinson velocity that bounds the propagation of quantum correlations in a many-body system \cite{Lieb1972,ourreview}.  Therefore, for times $t\ll L^\prime/v$, we can analyze the decay of the false vacuum by restricting to an $L^\prime \times L^\prime$ subregion surrounding $i$ where the QPC is obeyed (with high probability).  Indeed, we can identify the time scale $L^\prime/ v$ as a lower bound on the typical lifetime of the false vacuum as measured by any local observable.   Again, our new proof establishes that the slow decay of the false vacuum is not sensitive to the underlying theory being gapped.  This represents a step towards a more general theory of ``metastable manifolds" of low-energy states in many-body quantum systems; see \cite{Chen:2023idc,Bergamaschi:2025zlj} for a different perspective on this problem.

\emph{Outlook.}---
In summary, we have presented a stability proof using a quantum Peierls condition to establish the robustness of discrete group SSB in many-body quantum lattice models.  The result can, in principle, apply to gapless, frustrated, and long-range magnets. We also obtained bounds on the time scale for symmetry restoration which grows exponentially with system size in the quantum RBIM, and on the size of the symmetry-breaking field needed to destabilize a ``Schr\"odinger cat" ferromagnetic ground state or a false vacuum in dynamics.  Our results are consistent with physical expectations and place common intuition on a rigorous footing.

One intriguing open problem is the relationship between our proof of stable gapless phases based on quantum bottlenecks, and the conventional formalism for defining a stable phase of \emph{gapped} quantum matter based on quasiadiabatic continuity \cite{quasiadiabatic05,bachmann2012automorphic}.   In the context of the RBIM, the regions where the ``local" gap vanishes do not percolate, which means that as long as one is willing to tolerate a $\log(L)$-depth circuit, it may be possible to adiabatically connect the ground state of the gapped Ising model to the ground state of the RBIM.  At the same time, the many-body ground state in a gapless magnet is not stable, even as measured by local perturbations -- for example, a spin connected by weak bonds to its four nearest neighbors may flip its relative orientation in the ground state by adding perturbations that change the sign of the local $J_{ij}$. Also, the universality class of the transition out of the ferromagnet is believed to differ between the random and non-random Ising models.   While this transition is certainly outside of the regime of validity of our methods (which rely on being close to the ferromagnetic fixed point), it does raise the question of whether there is a meaningful difference between the gapped vs. gapless magnetic phases analyzed in this work.  Answering this question will be an important step towards developing a  mathematical classification for gapless quantum phases of matter.  

It is also \revise{an} interesting future direction to extend our method more broadly. Although our framework applies to quantum $H_0$ as quantum bottleneck theory \cite{q_bottle_Vedika}, we rely on $H_0$ being commuting so that there is a nice basis \eqref{eq:cleverbasis} to expand the Schr\"odinger equation. However, arguably more interesting gapless phases arise in fundamentally non-commuting Hamiltonians.  \revise{An important example is superfluidity, which has a gapless Goldstone boson.  It would be interesting to develop further techniques for the stability of superfluid phases.} 
Moreover, our current formulation requires a macroscopic energy barrier \eqref{eq:mainQPC} separating the low-energy sectors: A path connecting the two ground states with local updates necessarily involves a large domain wall at some middle stage. This thermal stability of $H_0$ from Peierls should by no means be necessary for ground state stability. As one can see from the 1d transverse-field Ising model, an $\order(1)$ energy barrier (gap in this case) can be already sufficient. 
With the energy barrier perspective, it may be also fruitful to extend our techniques to establish quantum many-body bottlenecks in open systems, following \cite{Gamarnik:2024hyk,q_bottle_Vedika}.  In particular, if our methods can be applied to $H_0$ representing the four-dimensional toric code \cite{topo_memory02}, it might be possible to establish the dynamical robustness of quantum memory to a broader class of perturbations than have thus far been studied.

\emph{Acknowledgements.}---
A.L. was supported by the National Science Foundation under CAREER Grant DMR-2145544.  C.Y. is partially supported by the National Science Foundation under Award No. 2016245  and by the Stanford Q-FARM Bloch Postdoctoral Fellowship in Quantum Science and Engineering.

\bibliography{biblio}

\comment{


\section*{Supplementary material}

## Appendix A: Classical statistical mechanics and the Peierls condition

This appendix will introduce the set of classical models of interest in our paper and review important known results about them.

### A.1. Peierls condition

**Definition A.1** (Configuration space). *Let $\Lambda$ denote the vertex set of a graph $G$ with $N = |\Lambda|$ a finite number. Place one $r$-dimensional bit at each vertex $i \in \Lambda$. The configuration space $\Omega$ corresponds to the set of all "bitstrings" $\mathbf{z}$: $\Omega = \mathbb{Z}_r^\Lambda$. The Hamming distance $\mathrm{Ham}(\mathbf{z}, \mathbf{z}')$ is the number of vertices where $\mathbf{z}, \mathbf{z}'$ differ, and*

$$\mathrm{Ham}(W, W') := \min_{\mathbf{z} \in W, \mathbf{z}' \in W'} \mathrm{Ham}(\mathbf{z}, \mathbf{z}') \tag{A1}$$

*for any $W, W' \subset \Omega$. For each subset $A \subset \Lambda$, define its edge boundary $\partial A$ as the set of edges in $G$ that connects one vertex in $A$ with one outside. Denote $\overline{\partial A}$ by the set of vertices in $A$ sharing some edge in $\partial A$.*

**Definition A.2** (Bottleneck and Peierls condition). *We say $(\{W_k, \Phi_k, \mathcal{B}_k\}_{k=1}^K, R)$ is a bottleneck structure if the following holds:*

1. *$W_k, \Phi_k \subset \Omega, R \geq 1$. Consider the "Hamming graph" with vertices $\Omega$ and edges connecting $\mathbf{z}, \mathbf{z}'$ if $\mathrm{Ham}(\mathbf{z}, \mathbf{z}') \leq R$. $\{W_k\}_{k=1}^K$ is pairwise disconnected, and $\Phi_k \subset \Omega \setminus W_k$ includes all vertices outside $W_k$ but connected to $W_k$ on this Hamming graph. We call each $W_k$ a well and $\Phi_k$ its bottleneck. Note that $\Phi_k$ may contain vertices disconnected to $W_k$, and may overlap with other $\Phi_{k'}, W_{k'}$.*

2. *$\mathcal{B}_k$ is a set where each bottleneck indicator $B \in \mathcal{B}_k$ is a function $B : \Omega \to \{0, 1\}$. For any $\mathbf{z} \in \Phi_k$, there exists at least one $B$ with $B(\mathbf{z}) = 1$.*

*We say Hamiltonian $H_0$ satisfies $(\Delta, \theta, L, R)$-Peierls condition (PC) for the bottleneck structure above, if the following conditions hold:*

1. *$\forall k$ and $\forall B \in \mathcal{B}_k$, there is an injective map $M_B : \Phi_k \to W_k \sqcup \Phi_k$ such that $\forall \mathbf{z} \in \Phi_k$ with $B(\mathbf{z}) = 1$,*

$$H_0(\mathbf{z}) \geq \Delta \cdot L + H_0(M_B \cdot \mathbf{z}). \tag{A2}$$

   *$M_B$ may depend on $k$ implicitly.*

2. *The total number of different bottleneck indicators $B \in \mathcal{B}_k$ is bounded: $\forall k$,*

$$|\mathcal{B}_k| \leq e^{\theta L}. \tag{A3}$$

Note that we assume a weaker version of Peierls condition comparing to the standard one (see e.g. Section 7.2 in [61]). Here we only need to constrain the energies of states in the bottleneck sets $\Phi_k$. For the example of studying the 2d Ising ferromagnetic phase, we only need to worry about the energy of states with system-size spanning domain walls of size $L \sim \sqrt{N}$. As a result, we will see that $H_0$ can be gapless while still satisfying (A2). An explicit example is the random bond Ising model, discussed in Appendix C.3.

### A.2. Peierls argument

**Definition A.3** (Thermal ensemble/Gibbs state). *The thermal ensemble/Gibbs state at inverse temperature $\beta$ is a probability distribution on $\Omega$ defined as follows:*

$$\mathbb{P}_\beta(\mathbf{z}) := \frac{e^{-\beta H_0(\mathbf{z})}}{\mathcal{Z}(\beta)} \tag{A4}$$

*where the partition function $Z$ and free energy $F(\beta)$ are defined as*

$$\mathcal{Z}(\beta) := e^{-\beta F(\beta)} := \sum_{\mathbf{z} \in \Omega} e^{-\beta H_0(\mathbf{z})}. \tag{A5}$$

Given a state subspace $W \subset \Omega$, we also define a restricted Gibbs state[1] as a probability distribution on $W$

$$\mathbb{P}_\beta^W(\mathbf{z}) := \frac{\mathrm{e}^{-\beta H_0(\mathbf{z})}}{\mathcal{Z}^W(\beta)} \mathbb{I}(\mathbf{z} \in W), \quad \text{where} \quad \mathcal{Z}^W(\beta) := \sum_{\mathbf{z} \in W} \mathrm{e}^{-\beta H_0(\mathbf{z})}. \tag{A6}$$

Here $\mathbb{I}(\cdot)$ returns 1 if $\cdot$ is true and 0 if it is false.

**Proposition A.4** (Peierls argument)**.** *If $H_0$ obeys a $(\Delta, \theta, L, R)$-Peierls condition (PC), then for*

$$\beta > \frac{\Delta}{\theta} \tag{A7}$$

*the Gibbs state has exponentially small support in the bottlenecks compared to the wells: $\forall k$,*

$$\mathbb{P}_\beta(\Phi_k) \le \frac{\mathrm{e}^{-(\beta\Delta-\theta)L}}{1 - \mathrm{e}^{-(\beta\Delta-\theta)L}} \mathbb{P}_\beta(W_k). \tag{A8}$$

*Proof.* This follows from a direct calculation:

$$\begin{aligned}
\sum_{\mathbf{z}\in\Phi_k} \mathrm{e}^{-\beta H_0(\mathbf{z})} &\le \sum_{B\in\mathcal{B}_k} \sum_{\mathbf{z}\in\Phi_k : B(\mathbf{z})=1} \mathrm{e}^{-\beta H_0(\mathbf{z})} \\
&\le \sum_{B\in\mathcal{B}_k} \sum_{\mathbf{z}\in\Phi_k : B(\mathbf{z})=1} \mathrm{e}^{-\beta[\Delta L + H_0(M_B \cdot \mathbf{z})]} & \text{(Using (A2))} \\
&\le \mathrm{e}^{-\beta\Delta L} \sum_{B\in\mathcal{B}_k} \sum_{\mathbf{z}' : \mathbf{z}' = M_B \cdot \mathbf{z} \text{ for some } \mathbf{z}\in\Phi_k} \mathrm{e}^{-\beta H_0(\mathbf{z}')} & (M_B \text{ is injective}) \\
&\le \mathrm{e}^{-\beta\Delta L} |\mathcal{B}_k| \sum_{\mathbf{z}' \in W_k \sqcup \Phi_k} \mathrm{e}^{-\beta H_0(\mathbf{z}')} & (M_B \text{ maps onto } W_k) \\
&\le \mathrm{e}^{-(\beta\Delta-\theta)L} \sum_{\mathbf{z}' \in W_k \sqcup \Phi_k} \mathrm{e}^{-\beta H_0(\mathbf{z}')} & \text{(Using (A3))}
\end{aligned}$$

Dividing both sides by $\mathcal{Z}(\beta)$ yields

$$\mathbb{P}_\beta(\Phi_k) \le \mathrm{e}^{-(\beta\Delta-\theta)L} \mathbb{P}_\beta(W_k \sqcup \Phi_k) = \mathrm{e}^{-(\beta\Delta-\theta)L} \left[\mathbb{P}_\beta(\Phi_k) + \mathbb{P}_\beta(W_k)\right], \tag{A9}$$

from which we can easily obtain (A8). □

The critical point about building a bottleneck is that when we introduce "dynamics" – either via a Gibbs sampler in classical statistical physics, or via quantum fluctuations when we turn to quantum models – the configuration space $\Omega$ becomes disconnected once the bottlenecks $\Phi_k$ are removed. At the same time, the bottleneck is very unlikely to be occupied in thermal equilibrium. For classical Markov chain dynamics, this is enough to show, by the following Proposition, that a Gibbs sampler will be trapped in only one well $W_k$ for a very long time [25], as was illustrated in Figure 1. The main result of this paper is that a similar result holds in quantum systems – we will build almost eigenstates of a many-body system with a bottleneck and prove that the system is trapped on one side of the bottleneck for extremely long times. For convenience, here we review the proof of a bottleneck bound in classical Markov chains:

**Proposition A.5.** *Consider a Markov chain transition matrix $T$ with entries obeying detailed balance*

$$T_{\mathbf{z}' \leftarrow \mathbf{z}} \mathbb{P}_\beta(\mathbf{z}) = T_{\mathbf{z} \leftarrow \mathbf{z}'} \mathbb{P}_\beta(\mathbf{z}'), \quad \forall \mathbf{z}, \mathbf{z}' \tag{A10}$$

*and locality*

$$\mathrm{Ham}(\mathbf{z}, \mathbf{z}') \ge R \quad \Rightarrow \quad T_{\mathbf{z}' \leftarrow \mathbf{z}} = 0. \tag{A11}$$

*For any $W = W_k$ in (A8), the restricted Gibbs state in that set is almost-steady:*

$$\left\| T\mathbb{P}_\beta^W - \mathbb{P}_\beta^W \right\|_1 := \sum_{\mathbf{z}\in\Omega} \left| \left(T\mathbb{P}_\beta^W\right)(\mathbf{z}) - \mathbb{P}_\beta^W(\mathbf{z}) \right| \le \frac{2\mathrm{e}^{-(\beta\Delta-\theta)L}}{1 - \mathrm{e}^{-(\beta\Delta-\theta)L}}. \tag{A12}$$

---

[1] The notion of restricted Gibbs state is sometimes known as "pure states" in spin glass literature; however, we do not use this terminology because it is easily confused with the notion of pure states in quantum mechanics.

*Proof.* Due to detailed balance (A10), probability distribution $T\mathbb{P}_\beta^W$ differs from $\mathbb{P}_\beta^W$ only by moving mass from some $\mathbf{z} \in W$ to some other $\mathbf{z}' \notin W$, which is contained in $\Phi := \Phi_k$ from (A11):

$$\begin{aligned}
\left\|T\mathbb{P}_\beta^W - \mathbb{P}_\beta^W\right\|_1 &= \sum_{\mathbf{z}\in\Phi, \mathbf{z}'\in W} T_{\mathbf{z}\leftarrow\mathbf{z}'}\mathbb{P}_\beta^W(\mathbf{z}') + \sum_{\mathbf{z}\in W, \mathbf{z}'} \left|T_{\mathbf{z}\leftarrow\mathbf{z}'}\mathbb{P}_\beta^W(\mathbf{z}') - \mathbb{P}_\beta^W(\mathbf{z})\right| \\
&= \sum_{\mathbf{z}\in\Phi, \mathbf{z}'\in W} T_{\mathbf{z}\leftarrow\mathbf{z}'}\mathbb{P}_\beta^W(\mathbf{z}') + \sum_{\mathbf{z}\in W, \mathbf{z}'} \left|(T_{\mathbf{z}'\leftarrow\mathbf{z}} - 1)\mathbb{P}_\beta^W(\mathbf{z})\right| \quad \text{(Detailed balance)} \\
&= 2 \sum_{\mathbf{z}\in W, \mathbf{z}'\in\Phi} T_{\mathbf{z}'\leftarrow\mathbf{z}}\mathbb{P}_\beta^W(\mathbf{z}) \\
&= 2\frac{\mathcal{Z}_\beta}{\mathcal{Z}_\beta^W} \sum_{\mathbf{z}\in W, \mathbf{z}'\in\Phi} T_{\mathbf{z}'\leftarrow\mathbf{z}}\mathbb{P}_\beta(\mathbf{z}) \\
&= 2\frac{\mathcal{Z}_\beta}{\mathcal{Z}_\beta^W} \sum_{\mathbf{z}\in W, \mathbf{z}'\in\Phi} T_{\mathbf{z}\leftarrow\mathbf{z}'}\mathbb{P}_\beta(\mathbf{z}') \quad \text{(Detailed balance)} \\
&\le 2\frac{\mathcal{Z}_\beta}{\mathcal{Z}_\beta^W} \sum_{\mathbf{z}'\in\Phi} \mathbb{P}_\beta(\mathbf{z}') \quad (\sum_\mathbf{z} T_{\mathbf{z}\leftarrow\mathbf{z}'} \le 1) \\
&\le 2\frac{\mathcal{Z}_\beta}{\mathcal{Z}_\beta^W} e^{-(\beta\Delta-\theta)L}\mathbb{P}_\beta(W_k) = \frac{2e^{-(\beta\Delta-\theta)L}}{1-e^{-(\beta\Delta-\theta)L}}, \quad (A13)
\end{aligned}$$

where the last step uses (A8). □

### A.3. Models with symmetries

We will mostly consider models with certain symmetries, with the Ising model as a motivating example:

**Example A.6** (Ising models). *The most important model we will study in this paper is the random bond Ising model. Given an undirected graph $(\Lambda, E)$ with a single connected component, we define*

$$H_0(\mathbf{z}) := \sum_{(i,j)\in E} J_{ij}\frac{1-z_iz_j}{2} =: \sum_{C_{ij}\in\Lambda_C} J_{ij}C_{ij} \quad (A14)$$

*Here $z_i = \pm 1$, $J_{ij}$ are bounded real numbers which may not all be positive. (A14) has a global $\mathbb{Z}_2$ symmetry:*

$$H_0(\mathbf{z}) = H_0(-\mathbf{z}), \quad (A15)$$

*with on-site symmetry generator $\ell : \ell \cdot z_i = -z_i$. The set $\Lambda_C$ of parity-checks are associated to the edges of the graph, which detect domain walls (DW)s between $+1$ and $-1$ regions.*

*In the standard Ising model, one takes $J_{xy} = J$ on every edge. It is straightforward to see that $H_0 = Jn$ for non-negative integer $n$ on each microstate. We say that $H_0$ is gapped because, no matter how large $|\Lambda|$ is, the difference between the two lowest values of $H_0$ remains finite.*

We must then establish that $H_0$ obeys a suitable Peierls-condition. Thankfully this has been done already for many models, including Ising models. For example, below we sketch the proof of the Peierls condition for the standard 2d Ising model, but the result extends to all higher dimensions as well.

**Proposition A.7** (Peierls condition for 2d Ising models). *Consider the Ising model (A14) with $J_{ij} = J > 0$ and interaction graph $(V, E)$ a two-dimensional $L_0 \times L_0$ square lattice with periodic boundary conditions. For any constant $R \ge 1$, this model obeys a $(J, \theta, c_R L_0, R)$-Peierls condition at sufficiently large $L_0$, where $\theta, c_R > 0$ are determined by $R$.*

*Proof.* We view the bits living on plaquettes and the checks living on links of the square lattice, so two links are connected if they share a common vertex. A loop is a set of connected links where each link is connected to exactly two other links. Any $\mathbf{z} \in \Omega$ can be determined by its domain wall (DW) configuration specifying which closed loops are excited, together with the value of the "net magnetization" (as each DW configuration corresponds to two distinct $\mathbf{z}$ related by symmetry). A more precise characterization of what "sign" to assign to each configuration will be discussed in the next paragraph for all low-energy configurations of interest. When 4 links sharing a common vertex are all

DWs, we choose the corresponding 2 loops to not intersect with each other and for the north/west and east/south edges to be joined together in loops.

We define wells $W_1 \sqcup W_2$ as those $\mathbf{z}$ whose DW configuration only has DW-loops $\lambda$ of bounded size

$$|\lambda| \leq L := \frac{L_0}{6R}. \tag{A16}$$

Since (A16) implies that no DW-loops can span across the whole system, there is one (and only one) connected region of plaquettes not separated by DW-loops that percolates across the torus. If $z_i = 1$ $(-1)$ in that region, we assign the corresponding $\mathbf{z}$ to $W_1$ $(W_2)$. In other words, bitstrings in $W_1$ contain a "sea" of $+1$s, with $-1$ regions forming *clusters* ("islands") inside of this sea.

The bottleneck space $\Phi_1 \cup \Phi_2$ is defined similarly to contain DW-loops $\lambda'$ of size

$$|\lambda'| \leq 4RL = \frac{2}{3}L_0. \tag{A17}$$

We further demand that there is at least one DW-loop of size $> L$ for any state in $\Phi_1 \cup \Phi_2$, so that the bottleneck space does not intersect with the wells. Since the DW-loops still cannot wind around the torus, $\Phi_1$ for example is still states with a $+1$-sea. Since flipping a single bit can join at most four DW-loops into one, any DW-loop $\lambda'$ of some $\mathbf{z}'$ within Hamming distance $R$ to $W_1$ includes at most $4R$ DW-loops of some $\mathbf{z} \in W_1$, so satisfies (A17). Therefore $\Phi_1$ includes all bitstrings within Hamming distance $R$ to $W_1$. The same holds for $\Phi_2, W_2$ by symmetry. Hence in this example, $\Phi_2 \cap \Phi_1 = \Phi_2 \cap W_1 = \emptyset$, etc.

For each loop $\lambda'$ of size

$$L < |\lambda'| \leq 4RL, \tag{A18}$$

we define a bottleneck indicator $B \in \mathcal{B}_1$ that returns 1 only when that loop is a DW-loop. We define the corresponding map $M_B$ as flipping all plaquettes enclosed by that loop. $M_B$ annihilates the DW-loop $\lambda'$ while keeping other DW-loops intact, so $M_B \cdot \mathbf{z} \in W_1 \sqcup \Phi_1$ if $\mathbf{z} \in \Phi_1$. Because $\lambda'$ is annihilated, (A2) holds with $\Delta = J$ from (A18) and (A14), since each DW link costs $J$ energy. We define exactly the same bottleneck indicators in $\mathcal{B}_2$ due to symmetry. We have therefore constructed a bottleneck structure $(\{W_k, \Phi_k, \mathcal{B}_k\}_{k=1}^2, R)$.

To establish PC, it remains to show (A3). Starting from a given link $E$, the number of loops of length $\ell$ containing $E$ is smaller than $3^{\ell-1}$ because each loop is a path on the links with at most 3 choices for choosing the next link (but not all of the enumerated paths are loops). Alternatively, the above enumeration of paths of length $4RL$ can also include loops of smaller length as we can allow for the path to return to its original link after fewer than $4RL$ steps. Hence we find that the total number of loops containing a given link is at most $3^{4RL-1}$. Since there are at most $2L_0^2$ links $E$ to choose to start the path,

$$|\mathcal{B}| \leq 2L_0^2 \, 3^{4RL-1} \leq 3^{RL} 3^{4RL} = e^{\theta L} \tag{A19}$$

from (A16) and $L_0 \geq 7$, where $\theta = 5R \log 3$. □

To generalize from the Ising model and formulate a class of models with on-site symmetries, we find it extremely helpful to borrow the terminology of checks from error-correcting codes. Note that in codes they are usually parity-checks, but here we do not require them to measure parity of bitstrings.

**Definition A.8** (Checks and syndromes). *A check is a function $C : \Omega \to \{0, 1\}$. We say $C$ is excited when $C(\mathbf{z}) = 1$. If $C(\mathbf{z})$ does not depend on $z_i$, we say $C$ does not act on $i$; otherwise $C$ acts on $i$. We choose a set of checks $\Lambda_\mathrm{C}$ such that each check acts on $\leq g$ ($g$ is a constant) vertices in $\Lambda$, and each $i \in \Lambda$ is acted on by $\leq g$ checks in $\Lambda_\mathrm{C}$.*

*For each $\mathbf{z}$, its syndrome $\mathbf{s} \in \mathbb{Z}_2^{\Lambda_\mathrm{C}}$ is a bitstring collecting the values of every $C(\mathbf{z})$. The set of allowed syndromes $\Omega_\mathrm{C} = \{\mathbf{s} : \mathbf{z} \in \Omega\}$ may not span the whole $\mathbb{Z}_2^{\Lambda_\mathrm{C}}$.*

This last point is crucial for the Ising model, which have redundancies in the set of checks for the above Peierls condition to hold. Indeed, we do not require a "minimal" set of checks $\Lambda_\mathrm{C}$ nor would that be desirable in finite-dimensional models.

**Remark A.9.** *In [5], a kind of Peierls condition with $L$ in (A2) replaced by $N$, known as linear confinement, for low-density parity-check (LDPC) codes was used to establish a kind of many-body localization, where all low-energy eigenstates are stable under perturbations. Actually, in this setting, it can be the case that $\Lambda_\mathrm{C}$ contains no redundant parity-checks, because LDPC codes are k-local models on expander graphs. In this paper we are most interested in systems where $L \ll N$, where the techniques of [5] do not directly apply.*

**Definition A.10** (On-site symmetry and symmetric Hamiltonians). *An on-site symmetry group $G$ is a set of invertible maps $\{\ell : \Omega \to \Omega\}$ that are on-site (i.e. $(\ell \cdot \mathbf{z})_i \equiv \ell_i \cdot z_i$ is determined by $z_i$ alone, where $\ell_i$ could depend on site $i$) and form a group with multiplication being the composition, and identity element being the identity map. We assume $G$ is a finite group in this paper.[2]*

*Each $\ell \in G$ and $A \subset \Lambda$ define a truncated on-site symmetry $\ell_A : \Omega \to \Omega$:*

$$\ell_A \cdot z_i = \begin{cases} \ell_i \cdot z_i, & i \in A, \\ z_i, & i \notin A. \end{cases} \tag{A20}$$

*Hamiltonian $H_0$ is symmetric under this symmetry group $G$ if and only if*

$$H_0(\mathbf{z}) = H_0(\ell \cdot \mathbf{z}), \quad \forall \ell \in G. \tag{A21}$$

We define checks for symmetric Hamiltonians as symmetric few-body operators, from which the bottleneck structure can be constructed.

**Definition A.11** (Checks and bottleneck for symmetric Hamiltonians). *Given an on-site symmetry group $G$, we say $(\{W_k, \Phi_k, \mathcal{B}_k\}_{k=1}^K, R)$ is a $G$-symmetric bottleneck structure, if it is a bottleneck structure in Definition A.2, and furthermore the following holds:*

1. *For any $k, k'$ there exists $\ell \in G$ such that $W_k = \ell \cdot W_{k'}$, $\Phi_k = \ell \cdot \Phi_{k'}$.*

2. *There exists a set $\Lambda_C$ of checks that are symmetric: $C(\ell \cdot \mathbf{z}) = C(\mathbf{z})$ ($\forall \ell \in G, \mathbf{z} \in \Omega$), from which the bottleneck indicators are built on: Each $B \in \mathcal{B}_k$ is of the form $B(\mathbf{z}) = C_1(\mathbf{z}) \cdot C_2(\mathbf{z}) \cdot ... \cdot C_{L_B}(\mathbf{z})$, checking whether some $L_B$ checks all return 1.*

This construction generalizes the Ising model example, where each $B$ is checking whether a loop is a DW-loop, i.e. whether the loop of checks are all excited.

To verify PC for a $G$-symmetric $H_0$ with the $G$-symmetric bottleneck structure above, one can usually choose $M_B = \ell_A$ in (A2) for some $\ell \in G$ and $A \subset \Lambda$ that de-excites all checks of $B$ to lower energy, where $B$ is "boundary" of $A$. The energy away from the boundary of $A$ is unchanged from symmetry. To have a large $L$, $B$ needs to be an "extended defect" so 1d Ising model for example does not satisfy PC. On the other hand, to verify (A3), it is sufficient to assume that each $B$ corresponds to a connected set of $L_B \sim L$ checks in a bounded-degree graph with checks living on vertices. This is because a connected cluster of size $L$ including a given vertex in a bounded-degree graph is at most $e^{cL}$ with constant $c$.

## Appendix B: Quantum bottlenecks from Peierls conditions

In this appendix, we are interested in studying the stability of a thermally stable $H_0$ with a Peierls condition in the presence of quantum fluctuations. Our setup begins by directly porting many of the definitions from Appendix A to quantum many-body systems:

**Definition B.1** (Quantum models with bottlenecks). *Consider the finite-dimensional Hilbert space $\mathcal{H} = \mathbb{C}[\Omega] = (\mathbb{C}^r)^{\otimes \Lambda}$, where one qudit is associated with each vertex in the set $\Lambda$. Alternatively, each basis vector of $\mathcal{H}$ corresponds to a microstate $\mathbf{z} \in \Omega$, with $\Omega$ the classical configuration space (Definition A.1). An operator $\mathcal{O}$ acting on $\mathcal{H}$ is $\kappa$-local if it is a superposition of operators of form $I_{A^c} \otimes \mathcal{O}'_A$ where $A \subset \Lambda$ obeys $|A| \leq \kappa$, and operator $\mathcal{O}'_A$ acts non-trivially only on qudits in $A$.*

*Define a set of checks $\Lambda_C$ as a commuting set of Hermitian operators where each $C \in \Lambda_C$ has two eigenvalues $0, 1$. Each check acts on $\leq g$ ($g$ is a constant) vertices in $\Lambda$, and each $i \in \Lambda$ is acted on by $\leq g$ checks in $\Lambda_C$. If a state $\psi$ has $C |\psi\rangle = 0$ ($|\psi\rangle$), then we say it has syndrome $0$ ($1$) on that check.*

*We say $(\{W_k, \Phi_k, \mathcal{B}_k\}_{k=1}^K, R)$ is a quantum bottleneck structure if the following holds:*

1. *$R \geq 1$, $W_k, \Phi_k$ are subspaces in $\mathcal{H}$. We use $W_k, \Phi_k$ to also represent the orthogonal projector onto those subspaces. $\{W_k\}_{k=1}^K$ is pairwise disconnected by any $R$-local operator $\mathcal{O}$:*

$$W_{k'} \mathcal{O} W_k = 0, \quad \forall k \neq k', \tag{B1}$$

   *and $\Phi_k \perp W_k$ includes all states of the form $\mathcal{O} |\psi\rangle$ with $R$-local $\mathcal{O}$ and $|\psi\rangle \in W_k$. We call each $W_k$ a well and $\Phi_k$ its bottleneck, with joint subspace $P_k := W_k + \Phi_k$. Note that $\Phi_k$ may overlap with other $\Phi_{k'}, W_{k'}$.*

---

[2] In all of the applications we will study in this paper, $G$ is a small group – in fact $G = \mathbb{Z}_2$ in all applications we will study in detail, which are Ising models and spin-offs thereof. It is straightforward to extend to other small on-site "flavor symmetries". However, all of the formalism that we are describing also applies to expander LDPC codes [5], where the finite group $G$ can be exponentially large in the number of degrees of freedom.

2. $\mathcal{B}_k$ is a set where each bottleneck indicator $B \in \mathcal{B}_k$ is an Hermitian operator acting on $\Phi_k$ with $B^2 = B$. There exists a basis $\{|\psi\rangle\}$ of $\Phi_k$ such that for each basis $|\psi\rangle$, there exists at least one $B$ with $B|\psi\rangle = |\psi\rangle$.

3. Each $B \in \mathcal{B}_k$ corresponds to a set of checks and $B = 1$ corresponds to those checks all being excited.[3] Let operator $E_B$ count the number of excited checks in those checks belonging to $B$ (so $L_B := \|E_B\|$ is the number of checks), and $Q_{B,n}$ projects onto the subspace for $E_B = n$. For any $|\psi\rangle = P_k |\psi\rangle$, $Q_{B,n} |\psi\rangle$ is still in subspace $P_k$. Denote $Q_{B,\geq n} = \sum_{n' \geq n} Q_{B,n'}$.

See Fig. 2 for an illustration. Note that although we have assumed a similar structure as symmetric models in Definition A.11 that each $B$ is a product of many checks,[4] the above definition applies generally beyond symmetric settings. We will see that this structure is crucial to decompose the Schrodinger equation into the different $E_B$ subspaces, from which we prove stability bounds.

**Definition B.2** (Quantum Peierls condition). *Let $H$, the Hamiltonian, be a Hermitian operator on $\mathcal{H}$. We say $H$ satisfies $(\Delta, \theta, L, R)$-quantum Peierls condition (QPC) for the above quantum bottleneck structure, if:*

1. *$H$ cannot bypass bottlenecks: $(I - P_k)H W_k = 0$, $\forall k$.*

2. *(A3) holds. $L_B \geq L$ for any $B \in \mathcal{B}_k$.*

3. *For any $B \in \mathcal{B}_k$ and $|\psi\rangle = P_k |\psi\rangle$ supported in subspace[5]*

$$Q_{B, \geq \frac{4}{5} L_B} |\psi\rangle = |\psi\rangle, \tag{B2}$$

*its energy satisfies*

$$\langle \psi | H | \psi \rangle \geq \Delta L_B + H_{k,\min}, \tag{B3}$$

*where*

$$H_{k,\min} = \min_{|\psi\rangle = P_k |\psi\rangle} \frac{\langle \psi | H | \psi \rangle}{\langle \psi | \psi \rangle} \tag{B4}$$

*is the lowest eigenvalue of $H$ restricted in $P_k$.*

Formally, QPC is not entirely comparable to PC Definition A.2: On one hand, QPC requires the extra structure of the bottleneck indicator, and constrains not just the state in bottleneck, but those triggering most checks of the bottleneck indicator. On the other hand, QPC could be weaker in the sense that (B3) compares energy with the ground state, instead of a particular state mapped from $\psi$ like (A2). Nevertheless, it seems to us in the examples we have studied that QPC follows from PC for the classical Hamiltonian $H_0$ even under perturbations: see Theorem C.3.

### B.1. Almost eigenstates

Before we begin to state our main results about the implications of QPC on the low-energy eigenstates of a many-body Hamiltonian, we will introduce the simple concept of almost eigenstates.[6]

**Definition B.3** (Almost eigenstate). *$|\psi\rangle$ is an $\delta$-almost eigenstate of Hamiltonian $H$ with eigenvalue $E$ if*

$$\|(H - E)|\psi\rangle\| \leq \delta \| |\psi\rangle \|. \tag{B5}$$

*Here and below, $\| |\psi\rangle \| = \sqrt{\langle \psi | \psi \rangle}$.*

$|\psi\rangle$ becomes an exact eigenstate when $\delta = 0$. For small $\delta$, $|\psi\rangle$ approximates an eigenstate in dynamics, i.e. it is almost stationary when evolved by $H$.

---

[3] It is *not* necessary that the violated checks in any $B$ are "connected". Although this is often the case, especially when we study Ising model $H_0$ (as is the main example throughout the paper), the sets $B$ will in general not be connected when e.g. applying this formalism to expander codes: see e.g. [5].

[4] In classical symmetric models, $W_k, \Phi_k, P_k$ naturally reduce to sets of bitstrings, while checks and bottleneck indicators reduce to classical functions on bitstrings.

[5] The constant $\frac{4}{5}$ is chosen for concreteness and can be adjusted. The factor $\Delta$ in (B3) can be chosen as $\frac{3}{5}$ times $\Delta$ in the classical PC (A2) for perturbed models; however, we do not want to carry around $\frac{3}{5}$ in every calculation!

[6] This object may go by the names "approximate eigenstate" or "pseudoeigenstate" in other literature.


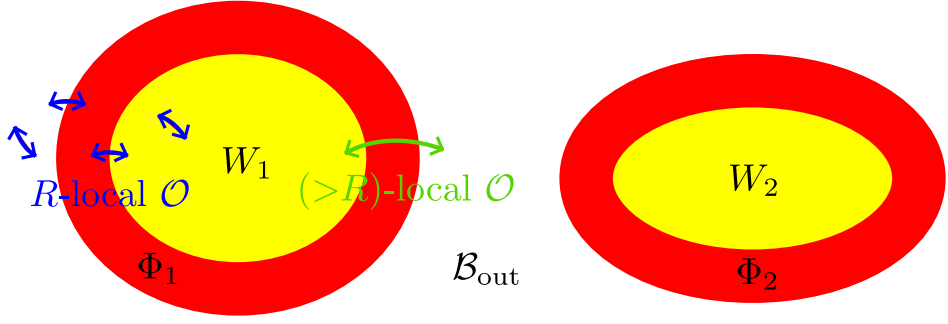

FIG. 2. Illustration of a quantum bottleneck structure. Here $\mathcal{B}_{\text{out}}$ is not required for the basic Definition B.1, but appears in the global version Definition B.7.

**Proposition B.4.** *If $|\psi\rangle$ is a $\delta$-almost eigenstate of $H$ with eigenvalue $E$, and $\langle\psi|\psi\rangle = 1$, then for any operator $A$ with Heisenberg evolution $A(t) = e^{itH} A e^{-itH}$,*

$$|\langle\psi|A(t)|\psi\rangle - \langle\psi|A|\psi\rangle| \leq 2\|A\|\delta t. \tag{B6}$$

*Proof.* Using the Duhamel identity we may write

$$e^{-iHt}|\psi\rangle = \left[|\psi\rangle + \int_0^t ds\, e^{-iHs}(H-E)e^{iEs}|\psi\rangle\right] e^{-iEt} =: [|\psi\rangle + |\phi\rangle] e^{-iEt}. \tag{B7}$$

Thus

$$\langle\psi(t)|A|\psi(t)\rangle - \langle\psi|A|\psi\rangle = \langle\phi|A|\psi\rangle + \langle\psi|A|\phi\rangle + \langle\phi|A|\phi\rangle \leq \||\phi\rangle\|_2 \|A\|(\||\psi\rangle\| + \||\psi\rangle + |\phi\rangle\|) \leq 2\|A\|\||\phi\rangle\|_2. \tag{B8}$$

Here and below, $\|A\|$ denotes the operator norm when acting on an oeprator rather than a vector. By the triangle inequality and the fact that a unitary operator has norm $\|e^{-i(H-E)s}\| = 1$,

$$\||\phi\rangle\|_2 \leq \int_0^t ds\, \left\|e^{-iHs}(H-E)e^{iEs}|\psi\rangle\right\|_2 \leq t\|(H-E)|\psi\rangle\|_2. \tag{B9}$$

Combining (B5) and (B9) we obtain (B6). $\square$

### B.2. Stability of ground states

Now we turn to tackle the main physics problem of interest in this paper: the structure of low-energy states in many-body quantum Hamiltonians subject to QPC.

**Definition B.5** (Perturbed Hamiltonian). *We study a quantum many-body system with Hamiltonian*

$$H = H_0 + V \tag{B10}$$

*where $H_0$ commutes with all checks: $[H_0, C] = 0$, and perturbation $V$ satisfies:*

1. *It is $q$-local with constant $q$, in the sense that $V = \sum_{F \subset \Lambda} V_F$ where $V_F$ is supported on $F$, and all $|F| \leq q$.*

2. *It has local strength*

$$\epsilon = \max_{i \in \Lambda} \sum_{F \ni i} \|V_F\|. \tag{B11}$$

In the literature, it is well-established [11] that so long as $H_0$ is short-range and gapped with a local topological order condition, for sufficiently small $\epsilon$ the Hamiltonian $H$ is also gapped and there is a quasilocal unitary that rotates the ground state subspace of $H_0$ to that of $H$. Although such stability results are expected to hold beyond gapped

systems, previous approaches [8–14] seem to rely fundamentally on the gap $\Delta$, which effectively block-diagonalize the two subspaces separated by the gap using small $\sim \epsilon/\Delta \ll 1$ local rotations.

The purpose of this paper is to prove weaker a notion of stability for an $H$ which obeys Definition B.2. Crucially, however, Definition B.2 allows for $H_0$ to be *long-range*, *gapless* and even *frustrated*. As far as we know, there is no quantum theory of stable phases of matter in the presence of these complications; though see [13, 14] for work on the stability of codes without spatial locality. Our main result, Theorem B.6, thus represents a first step towards a more general classification of phases of matter.

**Theorem B.6.** *Let $H$ be constructed as in Definition B.5. If $H$ satisfies $(\Delta, \theta, L, q)$-QPC and*

$$\epsilon = \zeta \cdot \frac{\Delta}{3g} e^{-\frac{5}{2}qg\theta} < \frac{\Delta}{3g} \times \min\left[\frac{1}{\sqrt{2}}, e^{-\frac{5}{2}qg\theta}\right], \tag{B12}$$

*with $0 < \zeta < 1$, then for any $k$ and arbitrary eigenstate $|\psi\rangle = P_k |\psi\rangle$ of the restricted Hamiltonian $P_k H P_k$:*

$$P_k H P_k |\psi\rangle = E|\psi\rangle, \tag{B13}$$

*with*

$$E \leq H_{k,\min} + \frac{1}{2}\Delta L \tag{B14}$$

*where*

$$H_{k,\min} := \min_{|\psi\rangle = P_k|\psi\rangle} \frac{\langle \psi | P_k H P_k | \psi \rangle}{\langle \psi | \psi \rangle}, \tag{B15}$$

*it has exponentially small support in bottleneck*

$$\|\Phi_k |\psi\rangle\|^2 \leq \exp\left[-L \cdot \frac{2}{5qg} \log \frac{1}{\zeta}\right]. \tag{B16}$$

*Furthermore, $|\psi\rangle$ is an $\delta$-almost eigenstate of $H$ with*

$$\delta = \|H\| \exp\left[-L \cdot \frac{1}{5qg} \log \frac{1}{\zeta}\right] \leq \exp\left[-L \cdot \frac{1}{6qg} \log \frac{1}{\zeta}\right], \tag{B17}$$

*if $\|H\| = \text{poly}(L)$ and $L$ is sufficiently large comparing to constants $q, g, \zeta$.*

We again emphasize that we need QPC for the perturbed Hamiltonian $H$, instead of the unperturbed one. This makes sense because e.g. symmetric $H_0$ should be stable only against symmetric perturbation $V$; asymmetric $V$ will violate QPC of $H$ (Example C.2).

*Proof.* We can apply the union bound to the classical probabilities $\langle \psi | B | \psi \rangle$ to find the system in each bottleneck $B \in \mathcal{B}_k$ to obtain:

$$\langle \psi | \Phi_k | \psi \rangle \leq \sum_{B \in \mathcal{B}_k} \langle \psi | B | \psi \rangle \leq e^{\theta L} \cdot \max_{B \in \mathcal{B}_k} \langle \psi | B | \psi \rangle = e^{\theta L} \cdot \max_{B \in \mathcal{B}_k} \|B |\psi\rangle\|^2. \tag{B18}$$

We then focus on one $B$ and bound $\|B|\psi\rangle\|^2$.

Following [5], we write

$$|\psi\rangle = \sum_{n=1}^{n_*+1} A_n |\psi_n\rangle \tag{B19}$$

where $A_n \geq 0$, and $|\psi_n\rangle$ is normalized and supported in different eigensubspaces of $E_B$. More precisely, $|\psi_1\rangle$ is supported in subspace $E_B = L_B$, $|\psi_2\rangle$ is supported in $E_B = L_B - 1, \cdots, L_B - qg$, $|\psi_3\rangle$ is supported in $E_B = L_B - qg - 1, \cdots, L_B - 2qg, \cdots$. By construction, each $|\psi_n\rangle$ is contained in subspace $P_k$. We choose

$$n_* = \left\lfloor \frac{L}{5qg} \right\rfloor + 1 \geq \frac{L}{5qg}, \tag{B20}$$

so that all $|\psi_n\rangle$ with $n \le n_*$ are supported in range $E_B \ge \frac{4}{5} L_B$; the remaining part of $|\psi\rangle$ corresponds to $|\psi_{n_*+1}\rangle$. Each $|\psi_n\rangle \propto \widetilde{Q}_n |\psi\rangle$ for some projector $\widetilde{Q}_n$, which is a sum over projectors $Q_{B,n'}$ with $n'$ in the corresponding $E_B$ window; we remind the reader that $[Q_{B,n'}, P_k] = 0$ from the assumptions in Definitions B.1 and B.5, which is why we may safely perform this decomposition.

Because each qudit is involved in at most $g$ checks and $V$ is $q$-local, it can change $E_B$ by at most $qg$. Furthermore, $H_0$ commutes with $E_B$, so expanding eigenstate equation (B13) using (B19) yields

$$\widetilde{Q}_n A_n P_k (H - E) |\psi_n\rangle = -\widetilde{Q}_n A_{n+1} P_k V |\psi_{n+1}\rangle - \widetilde{Q}_n A_{n-1} P_k V |\psi_{n-1}\rangle \tag{B21}$$

$\forall n = 1, \cdots, n_*$. Here and below we have set $A_0 = 0$. Taking inner product with $|\psi_n\rangle = \widetilde{Q}_n |\psi_n\rangle = P_k |\psi_n\rangle$,

$$A_n \langle \psi_n | (H - E) |\psi_n\rangle = A_{n+1} v_n + A_{n-1} \bar{v}_{n-1}. \tag{B22}$$

Hence,

$$A_n \frac{\Delta}{2} L_B \le A_{n+1} v_n + A_{n-1} \bar{v}_{n-1}, \tag{B23}$$

using QPC (B3) together with (B14). Here $v_n := -\langle \psi_n | V |\psi_{n+1}\rangle$ satisfies

$$|v_n| \le \left\| \sum_{F: F \text{ touches the } L_B \text{ checks}} V_F \right\| \le \sum_{i: i \text{ acted on by the } L_B \text{ checks}} \sum_{F: F \ni i} \|V_F\| \le g L_B \epsilon, \tag{B24}$$

because only those $V_F$ can change $E_B$, and those checks act on $\le gL_B$ qudits.

For $n = 1$, (B23) yields

$$A_1 \le \frac{2g\epsilon}{\Delta} A_2 \tag{B25}$$

using (B24). Assuming

$$A_{n-1} \le \frac{3g\epsilon}{\Delta} A_n \tag{B26}$$

holds (which indeed holds for $n = 2$), (B23) implies

$$A_n \frac{\Delta}{3} L_B \le A_n \left( \frac{\Delta}{2} L_B - \frac{3g\epsilon}{\Delta} g L_B \epsilon \right) \le A_n \frac{\Delta}{2} L_B - \bar{v}_{n-1} A_{n-1} \le g L_B \epsilon A_{n+1}, \tag{B27}$$

where we have used (B12) for the first inequality. (B27) thus implies (B26) for the next $n \to n+1$. By induction, (B26) holds for all $n \le n_* + 1$, so that

$$A_1 \le \frac{3g\epsilon}{\Delta} A_2 \le \cdots \le \left(\frac{3g\epsilon}{\Delta}\right)^{n_*} A_{n_*+1} \le \left(\frac{3g\epsilon}{\Delta}\right)^{\frac{L}{5qg}} \tag{B28}$$

using (B12), $A_{n_*+1} \le 1$ and (B20).

Since $\|B|\psi\rangle\| = A_1$, plugging (B28) into (B18) yields (B16):

$$\|\Phi_k |\psi\rangle\|^2 \le e^{\theta L} \left(\frac{3g\epsilon}{\Delta}\right)^{\frac{2L}{5qg}} = \left(\frac{3g\epsilon}{\Delta} e^{\frac{5}{2} qg\theta}\right)^{\frac{2L}{5qg}} \tag{B29}$$

using (A3). From (B12), we see that this is an exponentially small quantity in $L$.

Lastly, notice that

$$\|(H - E)|\psi\rangle\| = \|(I - P_k)(H - E)|\psi\rangle\| = \|(I - P_k) H \Phi_k |\psi\rangle\| \le \|H\| \cdot \|\Phi_k |\psi\rangle\|, \tag{B30}$$

where we have used $(I - P_k) H W_k = 0$. Combining (B16) and (B30) leads to (B17), where $\|H\|$ can be absorbed into the exponential at sufficiently large $L$ if itself is a polynomial of $L$. $\square$

The above construction of a $\delta$-almost eigenstate is specific to a finite-sized system. However, we will see in Appendix D how to use these methods to study the dynamics of local operators in many-body systems on infinite lattices.

### B.3. Global quantum Peierls condition and its consequence

**Definition B.7** (Global quantum Peierls condition). *We say that Hamiltonian $H$ satisfies $(\Delta, \theta, L, q)$-global QPC, if it satisfies $(\Delta, \theta, L, q)$ QPC with the following extra conditions:*

1. *The different subspaces do not overlap $P_k P_{k'} = 0$ ($k \neq k'$), and cannot be connected by $H$:*

$$P_k H P_{k'} = 0. \tag{B31}$$

   Let $P_{\text{out}} = I - \sum_{k=1}^{K} P_k$ project onto the subspace not included in any wells/bottlenecks.

2. *There exists a set $\mathcal{B}_{\text{out}}$ of operators $B^2 = B$, and a basis $\{|\psi\rangle\}$ of subspace $P_{\text{out}}$, such that for each basis $|\psi\rangle$, there exists at least one $B \in \mathcal{B}_{\text{out}}$ with $B|\psi\rangle = |\psi\rangle$.*

3. *Like point 3 in Definition B.1, each $B \in \mathcal{B}_{\text{out}}$ corresponds to a set of checks and $B = 1$ corresponds to those checks all being excited. Let operator $E_B$ count the number of excited checks in those checks belonging to $B$ (so $L_B := \|E_B\|$ is the number of checks), and $Q_{B, \geq n}$ projects onto the subspace for $E_B \geq n$. We require*

$$\langle \psi | H | \psi \rangle \geq \Delta L_B + H_{\min}, \tag{B32}$$

   *with $H_{\min}$ being the lowest eigenvalue of $H$, for any*

$$Q_{B, \geq \frac{4}{5} L_B} |\psi\rangle = |\psi\rangle. \tag{B33}$$

4. *$L_B \geq L$ for any $B \in \mathcal{B}_{\text{out}}$. The number of $B$s in $\mathcal{B}_{\text{out}}$ with $L_B = L'$ is smaller than or equal to $e^{\theta L'}$.*

*When this criterion is applied to classical Hamiltonians $H$ (which are diagonal in the basis labeled by the classical configuration space $\Omega$) we can refer to it as a classical global Peierls condition.*

This global QPC condition constrains the whole Hilbert space rather than just $P_k$s in QPC, where the energy estimate (B32) holds for any state in the $E_B$ window, not just those contained in some $P_k$. Interestingly, the QPC condition Definition B.2 can be formulated with this stronger version of energy estimate, which is already sufficient to show structures of global eigenstates (rather than eigenstates constrained in subspace $P_k$):

**Corollary B.8.** *Consider the same setting as Theorem B.6 where we further require (B3) to hold for any state (B2) not necessarily contained in $P_k$. Suppose $P_k$ commutes with all checks.[7] The following holds for any eigenstate $H|\psi\rangle = E|\psi\rangle$ of the global Hamiltonian with*

$$E \leq \frac{\Delta}{2} L + H_{\min}. \tag{B34}$$

*(B16) holds for any $k$. Moreover, for each well $W_k$, $P_k|\psi\rangle$ is either a $\delta$-almost eigenstate of $H$ with*

$$\delta = 2\sqrt{\|H\|} \exp\left[-\frac{1}{2}\left(\frac{L}{5qg} - 1\right) \log \frac{1}{\zeta}\right], \tag{B35}$$

*or very small $\|P_k|\psi\rangle\| \leq \delta$.*

*Proof.* We mimic the proof of Theorem B.6. Although $|\psi\rangle$ is no longer contained in $P_k$, we still have

$$\langle \psi | \Phi_k | \psi \rangle \leq \sum_{B \in \mathcal{B}_k} \langle \psi | P_k B P_k | \psi \rangle \leq \sum_{B \in \mathcal{B}_k} \langle \psi | B | \psi \rangle \leq e^{\theta L} \cdot \max_{B \in \mathcal{B}_k} \|B|\psi\rangle\|^2, \tag{B36}$$

following (B18), where we have used $[B, P_k] = 0$ so $P_k B P_k$ projects onto a smaller subspace than $B$. Focusing on one $B$, we have the same expansion (B19) where $\psi_n$ is no longer contained in a given $P_k$. Nevertheless, we do not need this information anymore because we have relaxed the range for which (B3) holds. Expanding $H|\psi\rangle = E|\psi\rangle$, we have (B21) with $P_k$ deleted in the equation. Taking inner product with $|\psi_n\rangle$ yields the same (B22). Using (B3)

---

[7] This holds trivially for classical models where the checks and $P_k$ are all classical operators.

and $H_{k,\min} \geq H_{\min}$ for bounding the energy, we obtain the same bound (B28) following the previous proof, which further leads to (B16) from (B36).

Since $(I - P_k)HW_k = 0$ and $H|\psi\rangle = E|\psi\rangle$,

$$\begin{aligned}(H - E)P_k|\psi\rangle &= (I - P_k + \Phi_k + W_k)(H - E)P_k|\psi\rangle \\ &= (I - P_k)(H - E)\Phi_k|\psi\rangle - (\Phi_k + W_k)(H - E)(I - P_k)|\psi\rangle \\ &= (I - P_k)(H - E)\Phi_k|\psi\rangle - \Phi_k(H - E)(I - P_k)|\psi\rangle.\end{aligned} \tag{B37}$$

The first term is easy to bound

$$\|(I - P_k)(H - E)\Phi_k|\psi\rangle\| \leq \|H - E\|\|\Phi_k|\psi\rangle\| \leq 2\|H\|\exp\left[-L \cdot \frac{1}{5qg}\log\frac{1}{\zeta}\right], \tag{B38}$$

using (B16). For the second term of (B37), replacing $|\psi\rangle$ in (B36) by $(H - E)(I - P_k)|\psi\rangle$ yields

$$\|\Phi_k(H - E)(I - P_k)|\psi\rangle\|^2 \leq e^{\theta L} \cdot \max_{B \in \mathcal{B}_k}\|B(H - E)(I - P_k)|\psi\rangle\|^2. \tag{B39}$$

Since

$$(I - P_k)|\psi\rangle = \sum_{n=1}^{n_*+1} A_n(I - P_k)|\psi_n\rangle, \tag{B40}$$

where $(I - P_k)|\psi_n\rangle$ is in the same $E_B$ window as $\psi_n$,

$$B(H - E)(I - P_k)|\psi\rangle = B(H - E)[A_1(I - P_k)|\psi_1\rangle + A_2(I - P_k)|\psi_2\rangle], \tag{B41}$$

because $H$ can only connect neighboring $n$ values. Taking norm of (B41),

$$\|B(H - E)(I - P_k)|\psi\rangle\| \leq \|H - E\|\sqrt{A_1^2 + A_2^2} \leq 2\|H\|\sqrt{\left(\frac{3g\epsilon}{\Delta}\right)^2 + 1}\left(\frac{3g\epsilon}{\Delta}\right)^{\frac{L}{5qg}-1} \leq \sqrt{6}\|H\|\left(\frac{3g\epsilon}{\Delta}\right)^{\frac{L}{5qg}-1} \tag{B42}$$

using (B28) and (B12) along with $\|H - E\| \leq 2\|H\|$.

Combining (B37), (B38), (B39) and (B42),

$$\|(H - E)P_k|\psi\rangle\| \leq (\sqrt{6} + \sqrt{2})\|H\|\exp\left[-\left(\frac{L}{5qg} - 1\right)\log\frac{1}{\zeta}\right] \leq \delta^2 \tag{B43}$$

using (B35). If $\|P_k|\psi\rangle\| \leq \delta$ then the proof is done; if not, $\|(H - E)P_k|\psi\rangle\| \leq \delta\|P_k|\psi\rangle\|$ so that $P_k|\psi\rangle$ is a $\delta$-almost eigenstate. □

In the above proof, we are not guaranteed that $|\psi\rangle$ has a $> \delta$ support in any $P_k$ space. To ensure this, we need the global QPC that there are no other wells overlooked:

**Corollary B.9.** *Under the same assumptions as Theorem B.6 with the QPC replaced by global QPC, for any eigenstate $H|\psi\rangle = E|\psi\rangle$ at low energy (B34), the conclusion of Corollary B.8 holds with a modified value*

$$\delta' = \sqrt{c_\zeta}\delta = c_\zeta \cdot 2\sqrt{\|H\|}\exp\left[-\frac{1}{2}\left(\frac{L}{5qg} - 1\right)\log\frac{1}{\zeta}\right] \tag{B44}$$

*where*

$$c_\zeta = \left(1 - \exp\left(-\frac{2}{5qg}\log\frac{1}{\zeta}\right)\right)^{-1/2}, \tag{B45}$$

*and (B16) is replaced by*

$$\|\Phi_k|\psi\rangle\|^2 \leq \max\left[1, \left(\frac{2c_\zeta\Delta\|H\|}{3g^2L\epsilon^2}\right)^2\right] \cdot \exp\left[-L \cdot \frac{2}{5qg}\log\frac{1}{\zeta}\right]. \tag{B46}$$

*Furthermore,*

$$\|P_{\text{out}}|\psi\rangle\| \leq c_\zeta\exp\left[-L \cdot \frac{1}{5qg}\log\frac{1}{\zeta}\right]. \tag{B47}$$

*Proof.* We restrict $H |\psi\rangle = E |\psi\rangle$ in subspace $P_k$ to obtain

$$P_k(H-E)P_k |\psi\rangle = -P_k(H-E)(I-P_k) |\psi\rangle = -P_k(H-E)P_{\text{out}} |\psi\rangle = -\Phi_k(H-E)P_{\text{out}} |\psi\rangle =: -|\chi\rangle \quad (B48)$$

using (B31). This returns to the previous case (B13) where $|\psi\rangle$ is replaced by $P_k |\psi\rangle$, except for the existence of an error term $|\chi\rangle$.

We first show that the error term is small: Following (B36),

$$\|P_{\text{out}} |\psi\rangle\|^2 \leq \sum_{B \in \mathcal{B}_{\text{out}}} \langle\psi|B|\psi\rangle \leq \sum_{L' \geq L} e^{\theta L'} \max_{B \in \mathcal{B}_{\text{out}}: L_B = L'} \|B |\psi\rangle\|^2 \quad (B49)$$

using point 4 of Definition B.7. Using the energy estimate (B32), the proof above that expands (B19) directly generalizes to bound

$$\|B |\psi\rangle\| \leq \left(\frac{3g\epsilon}{\Delta}\right)^{\frac{L_B}{5qg}}, \quad (B50)$$

so that (B49) becomes

$$\|P_{\text{out}} |\psi\rangle\|^2 \leq \sum_{L' \geq L} \exp\left[-L' \cdot \frac{2}{5qg} \log \frac{1}{\zeta}\right] = c_\zeta^2 \exp\left[-L \cdot \frac{2}{5qg} \log \frac{1}{\zeta}\right]. \quad (B51)$$

This bounds the error

$$\||\chi\rangle\| \leq \|H - E\| \|P_{\text{out}} |\psi\rangle\| \leq 2 \|H\| c_\zeta \exp\left[-L \cdot \frac{1}{5qg} \log \frac{1}{\zeta}\right] \leq c_\zeta \delta^2, \quad (B52)$$

where we have used (B35). As in the proof of Corollary B.8, this implies either $P_k |\psi\rangle$ is a $\sqrt{c_\zeta}\delta$-almost eigenstate, or $\|P_k |\psi\rangle\| \leq \sqrt{c_\zeta}\delta$.

To bound $\|\Phi_k |\psi\rangle\|$, we again use (B36) and expand $P_k |\psi\rangle = \sum_{n=1}^{n_*+1} A_n |\psi_n\rangle$ like (B19) for a particular $B \in \mathcal{B}_k$. (B48) then yields the same (B21) with

$$\widetilde{Q}_1 A_0 P_k V |\psi_0\rangle \mapsto \widetilde{Q}_1 |\chi\rangle. \quad (B53)$$

In other words, the iteration relation (B22) still holds but with initial condition $A_0 = 0$ replaced by

$$A_0 \leq \frac{\||\chi\rangle\|}{gL_B\epsilon}, \quad (B54)$$

and $v_0 = gL_B\epsilon$ satisfying (B24).

If the recursive assumption (B26) does not hold at $n = 1$, then

$$A_1 < \frac{\Delta}{3g\epsilon} A_0 \leq \frac{2c_\zeta \Delta \|H\|}{3g^2 L\epsilon^2} \exp\left[-L \cdot \frac{1}{5qg} \log \frac{1}{\zeta}\right] \quad (B55)$$

using (B52) and (B54). Otherwise, (B26) will hold iteratively and we get the previous bound (B28). In both cases,

$$\|\Phi_k |\psi\rangle\|^2 = A_1^2 \leq \max\left[1, \left(\frac{2c_\zeta \Delta \|H\|}{3g^2 L\epsilon^2}\right)^2\right] \cdot \exp\left[-L \cdot \frac{2}{5qg} \log \frac{1}{\zeta}\right] \quad (B56)$$

which completes the proof. $\square$

### Appendix C: Spontaneous symmetry breaking in quantum many-body systems

The main physical application of the formalism we have developed is to prove the existence and robustness of quantum phases of matter with spontaneous symmetry breaking. In the first subsection we will develop a general formalism and establish the circumstances under which we can guarantee a (global) QPC. Then we will discuss a number of non-trivial applications and extensions of the approach, using the two-dimensional Ising model as an illustration of the method (as we may use the same PC discussed in Proposition A.7).

### C.1. Locally symmetric Hamiltonians obey a quantum Peierls condition

Invoking the results of Appendix B requires knowledge that $H$ obeys a QPC. We now must establish circumstances under which the QPC holds. In this subsection, we will analyze the stability of symmetric $H_0$, discussed in Appendix A.3, perturbed by symmetric $V$. Actually for $q$-local $V$ that is long-range in terms of graph $G$, we require a stronger condition that it is locally symmetric:

**Definition C.1** (Locally symmetric perturbations). *Consider an on-site symmetry group $G$ represented on the Hilbert space by unitary operators $\{\ell\}$. Given a quantum bottleneck structure that is $G$-symmetric in the sense of satisfying Definition A.11 where the set of checks are symmetric, it defines a graph structure on $\Lambda$: two qudits are connected by an edge if they are acted on by one same check.*

*We say perturbation $V$ in Definition B.5 is locally symmetric, if there exists a decomposition $F = F_1 \sqcup F_2 \cdots \sqcup F_p$ such that each $F_j$ is connected on the graph $\Lambda$, $V_F = V_{F;F_1} \otimes V_{F;F_2} \otimes \cdots$ where each $V_{F;F_j}$ acts only on $F_j$, and*

$$\ell^\dagger V_{F;F_k} \ell = V_{F;F_k}, \quad \forall \ell \in G. \tag{C1}$$

**Example C.2.** *Symmetric $H_0$ can be unstable against symmetric but not locally symmetric $V$: Consider the example where $H_0$ is 2d Ising and $V = \frac{\epsilon}{N}\sum_{i<j} Z_i Z_j$ that satisfies (B11). One can verify that the state with all-zero on the left half and all-one on the right half has lower energy comparing to the unperturbed ground states [29, 62, 63].*

However, we now prove that a symmetric $H_0$ is robust against locally symmetric perturbations $V$, in the sense that the QPC is still obeyed.

**Theorem C.3.** *Consider a classical symmetric Hamiltonian $H_0$ with checks and bottleneck indicators given in Definition A.11 that satisfy $R \geq q$ and (A3). Suppose for every $B \in \mathcal{B}_k$, $L_B \geq L$ and there exists a truncated symmetry $\ell_A$ such that for each $\mathbf{z}$ with $E_B|\mathbf{z}\rangle = n|\mathbf{z}\rangle$ where $n \geq \frac{4}{5}L_B$,*

$$H_0(\mathbf{z}) \geq \Delta L_B + H_0(\ell_A \cdot \mathbf{z}), \tag{C2}$$

*and $\ell_A \cdot \mathbf{z}$ is in $P_k$ if $\mathbf{z}$ is in $P_k$. Furthermore, $|\partial A| \leq c_\partial L_B$ for some constant $c_\partial$ where $\partial A$ is the set of qudits in $A$ that are connected to the complement of $A$. Given the above assumptions, the perturbed $H = H_0 + V$ with any $q$-local locally symmetric $V$ satisfies $(\Delta - 2c_\partial \epsilon, \theta, L, R)$-QPC, where $\epsilon$ is given by (B11).*

*If $H_0$ furthermore satisfies global QPC in Definition B.7 where (B32) is replaced by (C2) for any $\mathbf{z}$ with $n \geq \frac{4}{5}L_B$, then $H = H_0 + V$ satisfies $(\Delta - 2c_\partial \epsilon, \theta, L, R)$-global QPC.*

*Proof of Theorem C.3.* We verify the criteria in Definition B.2 one-by-one. Since $H_0$ is classical, $(I - P_k)H_0 W_k = 0$. $(I - P_k)VW_k = 0$ because $R \geq q$. Therefore $(I - P_k)HW_k = 0$. Since (A3) and $L_B \geq L$ are assumed, we only need to verify (B3). Fix a given $B \in \mathcal{B}_k$, for any $|\psi\rangle$ in subspace (B2) with $P_k|\psi\rangle = |\psi\rangle$, $|\psi\rangle = \sum_\mathbf{z} a_\mathbf{z}|\mathbf{z}\rangle$ only contains $\mathbf{z}$ in $P_k$ satisfying $E_B|\mathbf{z}\rangle = n|\mathbf{z}\rangle$ with $n \geq \frac{4}{5}L_B$. (C2) implies

$$\langle\psi|H_0|\psi\rangle = \sum_\mathbf{z}|a_\mathbf{z}|^2 H_0(\mathbf{z}) \geq \sum_\mathbf{z}|a_\mathbf{z}|^2 [\Delta L_B + H_0(\ell_A \cdot \mathbf{z})] = \Delta L_B + \langle\psi|\ell_A^\dagger H_0 \ell_A|\psi\rangle. \tag{C3}$$

On the other hand,

$$\begin{aligned}
\langle\psi|\ell_A^\dagger V \ell_A|\psi\rangle - \langle\psi|V|\psi\rangle &= \langle\psi|\sum_{F:F\cap\partial A\neq\emptyset}\left(\ell_A^\dagger V_F \ell_A - V_F\right)|\psi\rangle \\
&\leq \sum_{i\in\partial A}\sum_{F\ni i}\left\|V_F - \ell_A^\dagger V_F \ell_A\right\| \\
&\leq \sum_{i\in\partial A}\sum_{F\ni i} 2\|V_F\| & \text{(Triangle inequality)} \\
&\leq |\partial A| \cdot 2\epsilon & \text{(Using (B11))} \\
&\leq 2c_\partial \epsilon L_B. & \text{(C4)}
\end{aligned}$$

Here the first line follows from $V$ being locally symmetric, because for any $F$ with $F \cap \partial A = \emptyset$, $V_F = \mathcal{O}_{F_\text{in}} \otimes \mathcal{O}_{F_\text{out}}$ where the two supports $F_\text{in}, F_\text{out}$ are contained in and outside $A$ respectively each being the union of some $F_k$, so that

$$\ell_A^\dagger V_F \ell_A = \ell_A^\dagger \mathcal{O}_{F_\text{in}} \ell_A \otimes \mathcal{O}_{F_\text{out}} = \ell^\dagger \mathcal{O}_{F_\text{in}} \ell \cdot \mathcal{O}_{F_\text{out}} = \mathcal{O}_{F_\text{in}} \otimes \mathcal{O}_{F_\text{out}} = V_F, \tag{C5}$$

using (C1). Combining (C3) and (C4),

$$\langle\psi| H |\psi\rangle - \langle\psi| \ell_A^\dagger H \ell_A |\psi\rangle = \langle\psi| H_0 |\psi\rangle - \langle\psi| \ell_A^\dagger H_0 \ell_A |\psi\rangle - \left( \langle\psi| \ell_A^\dagger V \ell_A |\psi\rangle - \langle\psi| V |\psi\rangle \right)$$
$$\geq \Delta L_B - 2c_\partial \epsilon L_B, \tag{C6}$$

which implies QPC (B3) with $\Delta \to \Delta - 2c_\partial \epsilon$ using variational principle $\langle\psi| \ell_A^\dagger H \ell_A |\psi\rangle \geq H_{k,\min}$, because $\ell_A |\psi\rangle = \sum_{\mathbf{z}} a_{\mathbf{z}} |\ell_A \cdot \mathbf{z}\rangle$ is still contained in $P_k$.

The above proof generalizes verbatim to the case of global QPC for $H_0$, where we do not keep track of which well $P_k$ the state is and use $\langle\psi| \ell_A^\dagger H \ell_A |\psi\rangle \geq H_{\min}$ in the end. □

Note that in Theorem C.3 we require a stronger condition than PC, namely (A2) holds not just for $\mathbf{z}$ in the bottleneck, but also for $\mathbf{z}$ "sufficiently close" to it measured by $E_B$. The Ising model is an example:

**Proposition C.4.** *The 2d Ising model in Proposition A.7 obeys a $(\Delta, \theta, L, R)$-global Peierls condition, and satisfies (C2) with $\Delta = 0.6J$ and $L = c_R L_0$.*

*Proof.* We supplement the bottleneck structure in Proposition A.7 by $\mathcal{B}_{\text{out}}$ where each $B \in \mathcal{B}_{\text{out}}$ detects whether a given loop of length $L_B \geq L$ is a DW-loop. The number of length-$L'$ loops is $\leq e^{\theta L'}$ due to the same path counting argument in Proposition A.7. Since any bitstring outside the two wells $P_1, P_2$ must have a large DW-loop, this yields a global bottleneck. To verify (C2), following the proof of Proposition A.7, $\ell_A \cdot \mathbf{z}$ always has $\leq 0.2 L_B$ checks excited in $\partial A$, so $H_0(\mathbf{z}) - H_0(\ell_A \cdot \mathbf{z}) \geq J(0.8 - 0.2)L_B \geq 0.6 J L_B$. □

### C.2. General results

**Definition C.5** (Classical spontaneous symmetry breaking). *Consider a classical configuration space $\Omega$ together with a group action with finite group $G$. Let $H_0 : \Omega \to \mathbb{R}$ be symmetric under $G$ and let $\mathbf{z}_k$ denote the ground states of $H_0$. The index $k$ lives in the coset space $G/G'$, where $G' \leq G$ is the stabilizer group of any $\mathbf{z}_k$. We say $H_0$ exhibits $G \to G'$ spontaneous symmetry breaking (SSB) if $G' \neq G$.*

*We say a $G$-symmetric bottleneck structure in Definition A.11 exhibits $G \to G'$ SSB, if the wells/bottlenecks $W_k/\Phi_k$ are labeled by cosets $k \in G/G'$ with $K = |G/G'|$, where the $G$-action on $\Omega$ descends to the left-multiplication action on the bottleneck sets: namely, $\ell \cdot (W_k, \Omega_k) = (W_{\ell \cdot k}, \Omega_{\ell \cdot k})$.*

*We say $H_0$ exhibits $G \to G'$ SSB with a $G$-symmetric bottleneck structure, if both $H_0$ and the bottleneck structure exhibits $G \to G'$ SSB where each well $W_k$ contains exactly one ground state $\mathbf{z}_k$, and $H_0$ satisfies the classical global PC in Definition B.7.*

Proposition A.5 demonstrates why the terminology of $G \to G'$ SSB is meaningful in the presence of the Peierls condition. Even though the systems under study are strictly finite dimensional, there is an almost steady state of any Markov chain restricted to a well labeled by $G/G'$. Indeed, from a practical perspective, we know that we observe SSB in finite-size experimental systems. Since some authors insist on taking the thermodynamic limit before defining SSB, we therefore wish to explicitly justify our terminology and why we find it appropriate to discuss SSB in large but finite systems. To do so, of course, we must provide reasonable definitions for SSB in quantum systems in the same spirit of Proposition A.5 in classical systems.

To begin, we find it helpful to introduce a somewhat formal notion of spontaneous symmetry breaking in a *state* in a quantum mechanical system. Let us stress that this definition does not rely on the existence of a Hamiltonian, only on the fact that the Hilbert space is a representation of $G$.

**Definition C.6** (Quantum state with spontaneous symmetry breaking). *Consider $\mathcal{H} = \mathbb{C}[\Omega]$, and assume there is a group action of $G$ on $\Omega$. This implies a permutation representation of $G$ on $\mathcal{H}$.*

*Let $|\psi\rangle \in \mathcal{H}$ be symmetric under $G$: for all $\ell \in G$, $\ell|\psi\rangle = |\psi\rangle$.[8] We say that this state has $(R, \delta, L_*)$-SSB from $G \to G'$ if there exists a bottleneck structure $(\{W_k, \Phi_k, \mathcal{B}_k\}, R)$ on the classical configuration space $\Omega$ with $G \to G'$ SSB, such that the different subspaces do not overlap, with $\text{Ham}(P_k, P_{k'}) > L_*$ if $k \neq k' \in G/G'$, and*

$$\left\| \left( I - \sum_{k \in G/G'} P_k \right) |\psi\rangle \right\| \leq \delta \||\psi\rangle\|. \tag{C7}$$

*In other words, most of the quantum state lies in the Hilbert space trapped in the wells.*

---

[8] Here we are using $\ell$ to refer to both the group element and its representation on $\mathcal{H}$, as the meaning is clear from context.

Here $L_*$ plays a role similar to $R$ in previous sections: We could have defined SSB of a quantum state without the extra parameter $L_*$, by replacing $P_k$ in (C7) by $W_k$ that are known to be $R$-apart. We have chosen $P_k$ in (C7) in order to directly apply our bound (B47) later. In practice, one can always shrinks the spaces $W_k$ (and $\Phi_k$ correspondingly) to insert distance $L_*$ between the disjoint spaces $P_k$s. In the 2d Ising Proposition C.4 for example, one can choose a smaller constant $c_R$ so that flipping $L_*/2$ bits in $P_1$ ($P_2$) cannot break the sea of all zeros (ones).

**Proposition C.7.** *Let $|\psi\rangle$ be a normalized state with $(R, \delta, L_*)$-SSB. Any $L_*$-local observable $A$ obeys*

$$\left| \langle\psi|A|\psi\rangle - \sum_{k\in G/G'} \langle\psi|P_k A P_k|\psi\rangle \right| < \delta \|A\| \tag{C8}$$

*Moreover, if $\ell^\dagger A \ell = A$ for all $\ell \in G$, then $\frac{\langle\psi|P_k A P_k|\psi\rangle}{\langle\psi|P_k|\psi\rangle}$ is independent of $k$, i.e. every "superselection sector" $P_k|\psi\rangle$ looks the same to symmetric local observables.*[9]

Following conventions in the literature, we could call the states $P_k|\psi\rangle$ (up to normalization) "superselection sectors". Proposition C.7, whose simple proof is omitted for brevity, tells us that $(R, \delta)$-SSB is sufficient to ensure that the symmetric state $|\psi\rangle$ is well-approximated by any of the superselection sectors $P_k|\psi\rangle$ to physical observables that are sufficiently local.

**Proposition C.8.** *Let $A_{\alpha,i}$ be a collection of $(L_*/2)$-local Hermitian observables: $i \in \tilde{\Lambda} \subseteq \Lambda$ denote vertices around which $A_{\alpha,i}$ acts locally, while $\alpha$ is a vector index in a non-trivial irreducible representation $\mathcal{U}$ of $G$ such that*

$$\ell^\dagger A_{\alpha,i} \ell = \sum_\beta \mathcal{U}_{\alpha\beta}(\ell) A_{\beta,i}. \tag{C9}$$

*If $|\tilde{\Lambda}|$ is proportional to $|\Lambda|$ with a constant prefactor, and a state $|\psi\rangle$ with $(R, \delta, L_*)$-SSB satisfies*

$$\left| \sum_{i\in\tilde{\Lambda}} \langle\psi_k|A_{\alpha,i}|\psi_k\rangle \right| > c|\tilde{\Lambda}| \tag{C10}$$

*for (at least) one well $|\psi_k\rangle = P_k|\psi\rangle$ and at least one $\alpha$ where $c$ is not too small*

$$c > 2\sqrt{\delta + 2\delta^2}\, \|A_{\alpha,i}\|, \tag{C11}$$

*then $|\psi\rangle$ has long-range order:*

$$\frac{1}{|\tilde{\Lambda}|^2} \sum_{i,j\in\tilde{\Lambda}} [\langle\psi|A_{\alpha,i} A_{\alpha,j}|\psi\rangle - \langle\psi|A_{\alpha,i}|\psi\rangle\langle\psi|A_{\alpha,j}|\psi\rangle] > \frac{1}{2} c^2. \tag{C12}$$

*Proof.* By the definitions above,

$$\sum_{i\in\tilde{\Lambda}} \langle\psi_{\ell\cdot k}|A_{\alpha,i}|\psi_{\ell\cdot k}\rangle = \sum_{i\in\tilde{\Lambda}} \langle\psi_k|\ell^\dagger A_{\alpha,i} \ell|\psi_k\rangle = \sum_{i\in\tilde{\Lambda}} \sum_\beta \mathcal{U}_{\alpha\beta}(\ell) \langle\psi_k|A_{\beta,i}|\psi_k\rangle \tag{C13}$$

where $|\psi_k\rangle = P_k|\psi\rangle$. Moreover since $|\psi\rangle$ has $(R, \delta, L_*)$-SSB and $A_{\alpha,i} A_{\alpha,j}$ acts on at most $L_*$ qudits by assumption, we see that for any $k \in G/G'$,

$$\left| \sum_{i\in\tilde{\Lambda}} \langle\psi|A_{\alpha,i}|\psi\rangle \right| \leq 2\delta|\tilde{\Lambda}|\|A_{\alpha,i}\| + \frac{1}{|G/G'|} \left| \sum_{\ell\in G} \sum_{i\in\tilde{\Lambda}} \langle\psi_{\ell\cdot k}|A_{\alpha,i}|\psi_{\ell\cdot k}\rangle \right|$$

$$\leq 2\delta|\tilde{\Lambda}|\|A_{\alpha,i}\| + \frac{1}{|G/G'|} \left| \sum_{\ell\in G} \sum_{i\in\tilde{\Lambda}} \sum_\beta \mathcal{U}_{\alpha\beta}(\ell) \langle\psi_k|A_{\beta,i}|\psi_k\rangle \right| = 2\delta|\tilde{\Lambda}|\|A_{\alpha,i}\| \tag{C14}$$

---

[9] A somewhat important semantic point is that we do take $|\psi\rangle$ itself to be symmetric under $G$ as otherwise we would say that the state has explicitly broken the symmetry. In the absence of a Hamiltonian this is the most natural way to distinguish between spontaneous and explicit symmetry breaking. So we would not say that superselection state $P_k|\psi\rangle$ has SSB.

where in the first line we peel off the part of $|\psi\rangle$ which is not trapped inside of the wells $P_k$, and in the second line we use the Great Orthogonality Theorem from representation theory: $\sum_{\ell \in G} \mathcal{U}_{\alpha\beta}(\ell) = 0$ for any non-trivial irreducible representation $\mathcal{U}$. At the same time, we do know that

$$\sum_{i,j \in \tilde{\Lambda}} \langle \psi | A_{\alpha,i} A_{\alpha,j} | \psi \rangle \geq -2\delta |\tilde{\Lambda}|^2 \|A_{\alpha,i}\|^2 + \left| \sum_{k \in G/G'} \sum_{i,j \in \tilde{\Lambda}} \langle \psi_k | A_{\alpha,i} A_{\alpha,j} | \psi_k \rangle \right|$$

$$\geq -2\delta |\tilde{\Lambda}|^2 \|A_{\alpha,i}\|^2 + \sum_{k \in G/G'} \left\langle \psi_k \left| \left( \sum_{i \in \tilde{\Lambda}} A_{\alpha,i} \right)^2 \right| \psi_k \right\rangle$$

$$\geq -2\delta |\tilde{\Lambda}|^2 \|A_{\alpha,i}\|^2 + \max_{k \in G/G'} \left| \left\langle \psi_k \left| \sum_{i \in \tilde{\Lambda}} A_{\alpha,i} \right| \psi_k \right\rangle \right|^2 \geq -2\delta |\tilde{\Lambda}|^2 \|A_{\alpha,i}\|^2 + c^2 |\tilde{\Lambda}|^2, \tag{C15}$$

where in the second line we used the fact that the operator being summed over is a positive semidefinite Hermitian operator to remove the absolute value bars, and in the third line we used (C10). Combining with (C13) and (C11) we obtain (C12). □

The two simple propositions above ensure that a state $|\psi\rangle$, formally defined in Definition C.6, satisfies all the criteria that a physicist would expect for a many-body state with SSB. With a meaningful criterion for how to discuss SSB in quantum states that involve a macroscopically large (but nevertheless finite) number of degrees of freedom, we may now go on to show that the QPC implies SSB: if the quantum ground state of $H_0$ has SSB, then $H_0 + V$ will also have SSB as long as $V$ is locally symmetric and sufficiently small.

**Corollary C.9** (Robustness of spontaneous symmetry breaking). *Consider the same setting as Theorem C.3 where $H_0$ satisfies $(\Delta + 2c_\partial \epsilon, \theta, L, q)$-global PC. Suppose $H_0$ exhibits $G \to G'$ SSB with that bottleneck structure. Then if $\epsilon$ obeys (B12), $H_0 + V$ obeys a $(\Delta, \theta, L, q)$-global QPC with a bottleneck structure with $G \to G'$ SSB. Moreover, every eigenstate of $H = H_0 + V$ ($H|\psi\rangle = E|\psi\rangle$) in the trivial representation of $G$ with $E$ bounded by (B34) has $(R, \delta, L_*)$-SSB with $\delta$ given by the right hand side of (B47), and $(R, L_*)$ inferred by the bottleneck structure.*

*Proof.* This is simply a restatement of previous results. $H_0 + V$ obeys a $(\Delta, \theta, L, q)$-global QPC by Theorem C.3. Corollary B.9 then implies that all low energy eigenstates are trapped in the wells $P_k$. Since the wells exhibit $G \to G'$ SSB (Definition C.5) and (C7) holds, the eigenstate $|\psi\rangle$ exhibits SSB (Definition C.6). □

Let us now confirm the physical expectation that a many-body system whose (unique) ground state on a thermodynamically large (but finite) lattice exhibits SSB, is extraordinarily sensitive to symmetry-breaking perturbations.

**Proposition C.10.** *Let $H$ be a Hamiltonian which obeys the assumptions of Corollary C.9 and whose unique ground state $|\psi\rangle$ is symmetric under $G$ and exhibits $G \to G'$ SSB given by (C8); let $P_k |\psi\rangle =: |\psi_k\rangle$. Suppose the ground state degeneracy is not huge:*

$$K \leq e^{c'_\zeta L} - 1, \quad \text{where} \quad c'_\zeta := \frac{1}{10qg} \log \frac{1}{\zeta}. \tag{C16}$$

*Let $A$ be an operator "charged" under the symmetry group $G$, with the property that for some $k_* \in G/G'$,[10]*

$$\langle \psi_k | A | \psi_k \rangle_{k \neq k_*} \geq \widehat{\epsilon} + \langle \psi_{k_*} | A | \psi_{k_*} \rangle, \tag{C17}$$

*where $\|A\| \leq \widehat{c} \widehat{\epsilon}$ for some constant $\widehat{c}$. Furthermore, $\langle \psi_k | A | \psi_{k'} \rangle = 0$ for any $k \neq k'$. Then, for any $\eta > 0$, the ground state $|\widehat{\psi}\rangle$ of the Hamiltonian $\widehat{H} := H + A$ obeys*

$$\langle \widehat{\psi} | \psi_{k_*} \rangle > 1 - \eta \tag{C18}$$

*so long as*

$$\widehat{\Delta} \geq c_1 \widehat{\epsilon} \geq c_1^2 \|H\| e^{-\frac{1}{4} c'_\zeta L} \tag{C19}$$

*for some constant $c_1$ determined by $q, g, \zeta, \eta$, where $\widehat{\Delta}$ is the spectral gap of $P_k H P_k$ (for any $k \in G/G'$) between its ground state and first excited state, and $L$ is sufficiently large.*

---

[10] If no such operator exists then one may define the state $|\psi\rangle$ as having local quantum topological order [15]. We are not aware of any model that has topological order as well as SSB of a global (0-form) symmetry, as we are studying in this paper. We doubt this can exist.

*Proof.* The energy $\widehat{E}$ of the new ground state is close to the unperturbed one: $\widehat{E} \leq H_{\min} + \|A\| \leq H_{\min} + \widehat{c}\,\widehat{\epsilon}$, so that

$$\langle \widehat{\psi}| H | \widehat{\psi}\rangle = \widehat{E} - \langle \widehat{\psi}| A | \widehat{\psi}\rangle \leq H_{\min} + 2\widehat{c}\,\widehat{\epsilon}. \tag{C20}$$

Let $Q_<$ project onto eigenstates of $H$ with energy $E \leq H_{\min} + \sqrt{\widehat{\Delta}\widehat{\epsilon}}$. From Markov inequality, (C20) implies

$$1 - \langle \widehat{\psi}| Q_< | \widehat{\psi}\rangle \leq \frac{\langle \widehat{\psi}| H | \widehat{\psi}\rangle - H_{\min}}{\sqrt{\widehat{\Delta}\widehat{\epsilon}}} \leq 2\widehat{c}\sqrt{\frac{\widehat{\epsilon}}{\widehat{\Delta}}}, \tag{C21}$$

i.e. $\widehat{\psi}$ is mainly supported in subspace $Q_<$.

We then show that $Q_<$ basically only contains ground states of each $P_k H P_k$. Following a strategy from [5], notice that given a $k$ and a global eigenstate $H|\phi\rangle = E_\phi |\phi\rangle$ with $\langle \phi|\phi\rangle = 1$,

$$\|P_k|\phi\rangle\| \leq \|(P_k H P_k - E_\phi)^{-1}\| \cdot \|H\| \cdot \|P_{\text{out}}|\phi\rangle\|. \tag{C22}$$

(C22) follows immediately from the projected Schrödinger equation $(P_k H P_k + P_k H P_{\text{out}})|\phi\rangle = E_\phi P_k |\phi\rangle$. By Corollary B.9, we see that $\|P_{\text{out}}|\phi\rangle\| \leq c_\zeta e^{-2c'_\zeta L}$, so this implies that

$$\|(P_k H P_k - E_\phi)^{-1}\| \geq \frac{\|P_k|\phi\rangle\|}{c_\zeta \|H\|} e^{2c'_\zeta L} \geq \frac{K^{-1/2}}{2c_\zeta \|H\|} e^{2c'_\zeta L} \geq \frac{1}{2c_\zeta \|H\|} e^{\frac{3}{2}c'_\zeta L} \tag{C23}$$

where we have used $\sum_k \|P_k|\phi\rangle\|^2 = 1 - \|P_k|\phi\rangle\|^2 \geq 1/4$ for sufficiently large $L$, chosen the $k$ that maximizes $\|P_k|\phi\rangle\|$, and used (C16) in the end. (C23) implies $E_\phi$ is within $2c_\zeta \|H\| e^{-\frac{3}{2}c'_\zeta L}$ distance to an eigenvalue of $P_k H P_k$. As a result, since each $P_k H P_k$ has gap $\widehat{\Delta}$ between the two lowest eigenstates, there is also a gap in the spectrum of $H$: For any $E_\phi \leq H_{\min} + \sqrt{\widehat{\Delta}\widehat{\epsilon}}$, it is exponentially close to the well ground state energy:

$$|H_{k,\min} - E_\phi| \leq 2c_\zeta \|H\| e^{-\frac{3}{2}c'_\zeta L}. \tag{C24}$$

Note that $H_{k,\min}$ is the same for different $k$ because the bottleneck structure has $G \to G'$ SSB.

Now we analyze the eigenstate equation per well (B48) for eigenstate $\phi$ satisfying (C24). If $|\chi\rangle = 0$, then $P_k|\phi\rangle \propto |\psi_k\rangle$. In general, the support (denoted by projector $Q_{k,>}$) of $P_k|\phi\rangle$ on the excited states of $P_k H P_k$ is small, because

$$\|(P_k H P_k - E_\phi)Q_{k,>} P_k|\phi\rangle\| \leq \||\chi\rangle\| + \|(H_{k,\min} - E_\phi)(I - Q_{k,>})P_k|\phi\rangle\|$$
$$\leq 4c_\zeta \|H\| e^{-\frac{3}{2}c'_\zeta L}, \tag{C25}$$

where we have used (B52) and (C24). So

$$\|Q_{k,>} P_k|\phi\rangle\| \leq \|Q_{k,>}(P_k H P_k - E_\phi)Q_{k,>}\|^{-1} 4c_\zeta \|H\| e^{-\frac{3}{2}c'_\zeta L} \leq \frac{2}{\widehat{\Delta}} 4c_\zeta \|H\| e^{-\frac{3}{2}c'_\zeta L}, \tag{C26}$$

because $E_\phi$ is at least $\frac{1}{2}\widehat{\Delta}$ far from the excited energies. The support of $|\phi\rangle$ outside $\mathrm{Span}(|\psi_k\rangle)$ (denoted by projector $Q_> = P_{\text{out}} + \sum_k Q_{k,>}$) is

$$\|Q_>|\phi\rangle\| \leq \left(\|P_{\text{out}}|\phi\rangle\|^2 + \sum_k \|Q_> P_k|\phi\rangle\|^2\right)^{1/2} \leq \sqrt{K+1}\,\frac{2}{\widehat{\Delta}} 4c_\zeta \|H\| e^{-\frac{3}{2}c'_\zeta L} \leq \frac{8}{\widehat{\Delta}} c_\zeta \|H\| e^{-c'_\zeta L} \tag{C27}$$

using (C16). Since $\sum_\phi \|(I - Q_>)|\phi\rangle\|^2 \leq \mathrm{Tr}(I - Q_>) \leq K$, and each $\|(I - Q_>)|\phi\rangle\| \geq 1/2$ from (C27), there are at most $2K$ eigenstates $\phi$ in range (C24). Furthermore, $Q_<|\widehat{\psi}\rangle$ is supported in those eigenstates, so

$$\left\|Q_> Q_< |\widehat{\psi}\rangle\right\| \leq \sqrt{2K} \max_\phi \|Q_>|\phi\rangle\| \leq \frac{8\sqrt{2}}{\widehat{\Delta}} c_\zeta \|H\| e^{-\frac{1}{2}c'_\zeta L} \tag{C28}$$

using (C16).

We now evaluate energy of $|\widehat{\psi}\rangle = \sum_k \alpha_k |\psi_k\rangle + \cdots$ and consider variation in $\alpha_k$:

$$\langle \widehat{\psi}| \widehat{H} | \widehat{\psi}\rangle = \langle \widehat{\psi}| (I - Q_<)\widehat{H}(I - Q_<)|\widehat{\psi}\rangle + \langle \widehat{\psi}| Q_< \widehat{H} Q_< |\widehat{\psi}\rangle + \epsilon_<$$
$$= \langle \widehat{\psi}| (I - Q_<)\widehat{H}(I - Q_<)|\widehat{\psi}\rangle + \sum_k |\alpha_k|^2 \langle \psi_k|A|\psi_k\rangle + \epsilon_> + \epsilon_<, \tag{C29}$$

where

$$|\epsilon_<| \leq 2\left|\langle\widehat{\psi}|Q_<\widehat{H}(I-Q_<)|\widehat{\psi}\rangle\right| \leq 2\|A\|\left\|(I-Q_<)|\widehat{\psi}\rangle\right\| \leq 2\widehat{c}\widehat{\epsilon}\left(2\widehat{c}\sqrt{\frac{\widehat{\epsilon}}{\widehat{\Delta}}}\right)^{1/2} \tag{C30}$$

using (C21), and

$$|\epsilon_>| \leq 2\left|\sum_k \alpha_k^* \langle\psi_k|\widehat{H}Q_>Q_<|\widehat{\psi}\rangle\right| \leq 3\|H\|\left\|Q_>Q_<|\widehat{\psi}\rangle\right\| \leq \frac{24\sqrt{2}}{\widehat{\Delta}}c_\zeta\|H\|^2 e^{-\frac{1}{2}c'_\zeta L}. \tag{C31}$$

Therefore,

$$|\epsilon_> + \epsilon_<| \leq c_2\widehat{\epsilon} \tag{C32}$$

with constant $c_2$ determined by $c_1$ in (C19) and $c_2 \to 0$ $(c_1 \to \infty)$. When varying $\alpha_k$, the change of $\epsilon_> + \epsilon_<$ is therefore subdominant comparing to the change in $\sum_k |\alpha_k|^2 \langle\psi_k|A|\psi_k\rangle$, where by putting all mass in $k_*$ gains energy $\widehat{\epsilon}$ from (C17). Since the first term in (C29) does not depend on $\alpha_k$, (C29) is minimized at $|\alpha_{k_*}|^2 \geq 1 - c_2$. This implies that for any $\eta > 0$, one can choose a sufficiently large constant $c_1$ in (C19) such that (C18) holds. □

In Proposition C.12 we show that an $A$ satisfying the assumptions of Proposition C.10 exists in 2d Ising models. The argument there is expected to hold more generally, although we did not construct a detailed bottleneck structure for more general models.

### C.3. Random bond Ising models

**Theorem C.11.** *Consider the 2d Ising model $H_0$ in Proposition A.7 where $J_{ij}$ are instead iid random variables drawn from a probability density function $f(J)$ with compact support in the domain $[-J_1, J_2]$. Suppose that*

$$\mathbb{P}[J_{ij} > \Delta'] = 1 - p \tag{C33}$$

*and that the inequalities $0 \leq J_1 < \Delta < \Delta' < J_2$ hold. Then if*

$$\chi := a\log\frac{a}{p} - a + p > \theta \tag{C34}$$

*where*

$$a := \frac{4\Delta' - 5\Delta - J_2}{5(\Delta' + J_1)} > p, \tag{C35}$$

$H_0$ *has a $(\Delta, \theta, L, R)$-global QPC with (C2), with very high probability at large $L = c_R L_0$ ($c_R$ is a constant determined by $R$ in Proposition A.7):*

$$\mathbb{P}(\text{global QPC is violated}) \leq \frac{e^{-(\chi-\theta)L}}{1 - e^{-(\chi-\theta)}}. \tag{C36}$$

*As a result, under small symmetric perturbations, $H_0 + V$ has $\mathbb{Z}_2 \to 0$ SSB at low energies as Corollary C.9.*

Note that as long as $4\Delta' > J_2$, condition (C34) always holds with some constant $\Delta > 0$ at sufficiently small $p$.

*Proof.* We use the same global bottleneck structure of the 2d Ising model in Proposition C.4. The goal is to verify that (C2) holds with high probability. Let us first fix a specific bottleneck $B$, i.e. a length-$L_B$ loop, that involves checks $C_1, \cdots, C_{L_B}$. We sort the coupling coefficients as $J_{C_1} \leq J_{C_2} \leq \cdots \leq J_{C_{L_B}}$. For (C2) to hold for all states with $\geq \frac{4}{5}L_B$ checks satisfied, it suffices to consider the worst case where the satisfied ones are the smallest $J$s:

$$H_0(\mathbf{z}) - H_0(\ell_A \cdot \mathbf{z}) \geq E_0 := \sum_{\mu=1}^{\frac{4}{5}L_B} J_{C_a} - \sum_{\mu > \frac{4}{5}L_B}^{L_B} J_{C_a}. \tag{C37}$$

A simple (not tight) bound on this comes from the following simple argument. If $M$ of the $J_C$s are $< \Delta'$, the smallest possible value of $E_0$ is

$$E_0 \geq \left(\frac{4}{5}L_B - M\right)\Delta' - MJ_1 - \frac{1}{5}L_B J_2. \tag{C38}$$

If we want $E_0 \geq \Delta L_B$ then we need

$$\frac{M}{L_B} \leq \frac{4\Delta' - 5\Delta - J_2}{5(\Delta' + J_1)} = a. \tag{C39}$$

The probability that this condition is violated is given by a Chernoff bound:

$$\mathbb{P}\left(M > \frac{4\Delta' - 5\Delta - J_2}{5(\Delta' + J_1)}L_B\right) \leq \left(\frac{e^\delta}{(1+\delta)^{1+\delta}}\right)^{pL_B} = e^{-\chi L_B} \tag{C40}$$

where $\delta = \frac{a}{p} - 1 > 0$. From (C34), (C40) is exponentially small in $L_B$. Furthermore, the union bound implies that the probability that we encounter an energy barrier violating the $(\Delta, \theta, L, R)$ global Peierls condition at scale $L'$ is smaller than

$$\mathbb{P}(\text{violation anywhere at scale } L') \leq e^{(\theta - \chi)L'}. \tag{C41}$$

Since $\theta < \chi$, we can sum over all scales $L'$ to find a negligible probability that the global QPC is violated anywhere with another union bound, leading to (C36).

The stability of SSB follows from Corollary C.9. □

The method used to prove the theorem above can be readily generalized to establish the robustness of the global QPC in any other system where an $H_0$ obeying a global QPC is modified such that the couplings of $H_0$ are random with a sufficiently strong bias, as in (C33).

### C.4. Order parameter and explicit symmetry breaking in the 2d Ising model

In the discussion of Proposition C.10, we assumed the existence of a charged operator $A$ that could detect which "well" a wave function belonged to. On physical grounds it is quite obvious what this operator $A$ ought to be. For example, in a two-dimensional magnet with $\mathbb{Z}_2 \to 0$ SSB, we would like to take

$$A := \frac{1}{|\Lambda|}\sum_{i \in \Lambda} Z_i \tag{C42}$$

to be the average magnetization. As $Z_i$ is charged under the broken symmetry, $A$ is called an **order parameter**.

Unfortunately, the well $P_1$ with a 0-sea does *not* restrict to the positive magnetization sector as we have so far constructed them. For example, in this same Ising model, we could consider a configuration that has a large density of packed bubbles of size $D$, each of which has a strictly finite size; the number of flipped spins could scale as $\sim (\frac{L_0}{D+2})^2 D^2$ while the length of the domain wall scales as $(\frac{L_0}{D+2})^2 4D$. This means that the energy density of this configuration is small compared to $\epsilon$, there are no large domain walls, yet the magnetization is still opposite sign to most states in the well. Hence we would like to establish a bottleneck/QPC criterion that demonstrates this configuration is also unlikely to be observed.

**Proposition C.12.** *The 2d Ising model in Proposition A.7 obeys a $(\Delta, \theta, L, R)$-global Peierls condition with the two wells distinguishable with order parameter* (C42):

$$P_1 A P_1 > \frac{1}{4}, \quad P_2 A P_2 < -\frac{1}{4}. \tag{C43}$$

*The bottleneck structure underlying the Peierls condition obeys* (C2) *with $\Delta = 0.6J$ and $L = c_R L_0$ for sufficiently large $L_0$.*

*Proof.* In Proposition C.4, we obtained a global bottleneck structure with wells denoted here by $W'_k, P'_k$ ($k = 1, 2$) with bottleneck indicators $\mathcal{B}'_k$, and a set of bottleneck indicators $\mathcal{B}'_{\text{out}}$ for states outside $P'_k$s. Here we construct a new global bottleneck structure by deleting some states in the $P'_k$s so that the new $P_k$ is a subspace of $P'_k$. We do this by appending new bottleneck indicators $B$ that would detect any bitstring with small polarization

$$A \in \left[-\frac{1}{3}, \frac{1}{3}\right]. \tag{C44}$$

The new well $W_k$ is then bitstrings in $W'_k$ that is not detected by any of the new $B$. If a bitstring in $P'_k$ is detected by some new $B$, we put it in either the new bottleneck space $\Phi_k$ (if it is of $R$-Hamming-distance to $W_k$), or the new outer space. The bitstrings in $P'_k$ not detected by the new $B$s are still in $P_k$.

We first show that for each $\mathbf{z}$ satisfying (C44), there exists some new bottleneck indicator $B$ that detects it $B|\mathbf{z}\rangle = |\mathbf{z}\rangle$, such that the total number of such $B$s (with a fixed size) is still bounded by $e^{\theta L'}$. Here each $B$ detects whether a particular set of loops are DW-loops, with $L'$ being the total length of the loops.

Observe that for any $\mathbf{z}$ with small polarization (C44), there is at least one row of bits in the $L_0 \times L_0$ lattice with $\geq \frac{1}{3}L_0$ 0s and $\geq \frac{1}{3}L_0$ 1s. Let's keep track of all the DW-loops that pass through this row. Notice that for $\mathbf{z} \in P'_k$, every DW loop that passes through this row is contractible, since the original bottleneck structure forbids any non-contractible loop whose length is at least $L$. Apparently their total length

$$L' \geq \frac{1}{3}L_0, \tag{C45}$$

so that the $B$ detecting this set of loops detects the bitstring. See Fig. 3(a) for an illustration. We then count the number $N(L')$ of such $B$s of fixed total length $L'$. If the places where DWs intersect the row of interest are held fixed (we will later sum over all possible locations), consider enumerating all the ways to close these DWs outside of the specified row. Note that each closed DW would have length $d_j$, with the total length

$$\sum_{j=1}^{p} d_j \leq L'. \tag{C46}$$

Here $p$ is the number of DWs in the row: see Figure 3. The number of ways to grow the paths is bounded by

$$\sum_{\{d_j > 0\}: \sum_{j=1}^{p} d_j \leq L'} \prod_{j=1}^{p} 3^{d_j} \leq 3^{L'} \times \left|\left\{\{d_j > 0\}: \sum_{j=1}^{p} d_j \leq L'\right\}\right| \leq 3^{L'} \times 2^{L'} \tag{C47}$$

because there are 3 choices to grow a path further by length 1, as in the proof of Proposition A.7, and one can interpret each sequence $(d_1, d_2, \cdots)$ as a length-$L'$ bitstring with $d_1 - 1$ zeros followed by 1, then $d_2 - 1$ zeros followed by 1 ... and finally 0s.

The number of DW-loop configurations with those DWs in the row fixed is upper bounded by the number of such paths (C47). Since there are $2^{L_0}$ ways to choose the fixed DWs in a given row, and there are $L_0$ rows in total,

$$N(L') \leq L_0 \times 2^{L_0} \times 6^{L'} \leq e^{\theta' L'} \tag{C48}$$

using (C45) for some constant $\theta'$. Therefore we have constructed a new global bottleneck structure where

$$W_1 A W_1 > \frac{1}{3}, \quad W_2 A W_2 < -\frac{1}{3}, \tag{C49}$$

and at the same time the number of bottleneck indicators of size $L'$ is bounded by $e^{\theta L'}$ for some constant $\theta$. Here $\theta > \theta'$ because beyond (C48) there are also old bottleneck indicators in $\mathcal{B}'_k, \mathcal{B}'_{\text{out}}$. (C49) implies (C43) because we put all bitstrings with (C44) that are $R$-Hamming-distance far from $W_1$ out of $P_k$s, which includes all states with $A \in \left[-\frac{1}{4}, \frac{1}{4}\right]$ due to their extensive polarization difference with (C49).

To verify (C2) for $H_0$, we only need to check the new bottleneck indicator $B$ with total length $L_B$. Since $B$ is a set of DW-loops $B_1, B_2, \cdots$ with interiors $A_1, A_2, \cdots$, we let $A = A_1 + A_2 + \cdots$[11] such that $\ell_A$ flips the bits surrounded by the loops. Since here we are just flipping possibly multiple regions, $|\partial A| \leq c_\partial L_B$ and (C2) still hold. $\square$

---

[11] Here each set $A$ is interpreted as a bitstring and $+$ is the binary sum.

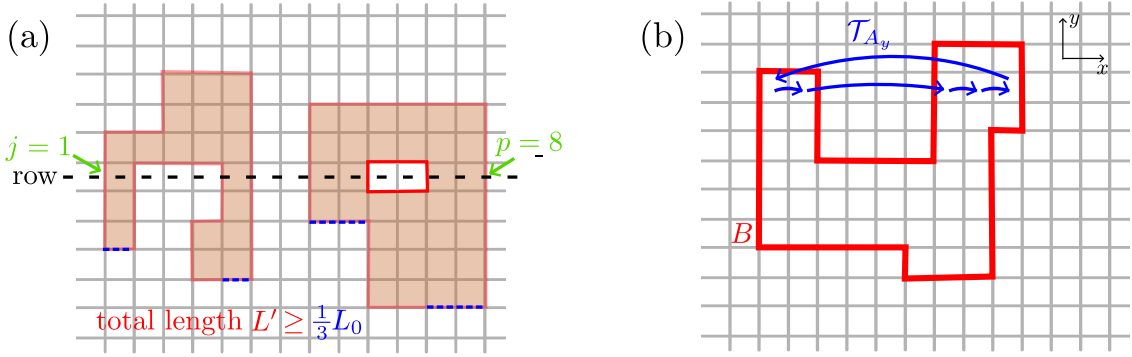

FIG. 3. (a) Illustration of a new bottleneck indicator in the proof of Proposition C.12 that detects $\geq \frac{1}{3}L_0$ 0s and $\frac{1}{3}L_0$ 1s in a row, based on the DW-loops (red) that cross the row. Although the number of crossings $p$ may be small, the total length $L' \geq \frac{1}{3}L_0$ of the DW-loops has to be large. (b) In the proof of Theorem C.14, we defined a truncated translation operator $\mathcal{T}_A$ along the $x$ direction of a region $A$ corresponding to a bottleneck indicator $B$ (red loop).

**Corollary C.13.** *Let $A$ be the order parameter in (C42). Consider the random field Ising model in Theorem C.11 under perturbation $V + \widehat{\epsilon} A$ where $V$ is symmetric with local strength $\epsilon + \widehat{\epsilon} \leq c_1$, and*

$$\widehat{\epsilon} \geq e^{-c_2 L_0}, \tag{C50}$$

*for some constants $c_1, c_2, c_3 > 0$. The ground state lies in well $P_2$ satisfying (C43), with leakage outside $P_2$ bounded by $e^{-c_3 L_0}$.*

*Proof.* This result directly follows from Theorem C.3, Proposition C.12, and Theorem C.11. In particular, comparing to Proposition C.10, we do not require a gap $\widehat{\Delta}$ in the wells since we know the whole well is distinguishable from $A$; the result just comes from a variational ansatz $|\psi\rangle = |\psi_1\rangle + |\psi_2\rangle + |\psi_{\text{out}}\rangle$ for the two wells and noticing that the optimal ansatz is to gain energy from $\widehat{\epsilon}$ by making $|\psi_1\rangle$ small. $\square$

This result proves the standard lore that SSB states can be constructed by finding the ground state of a model in an "infinitesimal" external field and taking the limit $L \to \infty$ before $\widehat{\epsilon} \to 0$. We see that this construction remains valid even when the phase is gapless and frustrated, and characterize just how tiny the symmetry breaking parameter $\epsilon$ needs to be to cause the vacuum (ground state) to lie in a single symmetry sector.

### C.5. Ising model in an antiferromagnetic longitudinal field

For simplicity, we have formulated our theory of robust SSB for on-site symmetries. Here, we show that our method can apply to more general models, with a little bit of additional work. Consider the 2d Ising model $H_0$ in Proposition A.7 with a longitudinal field perturbation

$$V = h \sum_{i \in \Lambda_e} Z_i - h \sum_{i \in \Lambda_o} Z_i + \cdots \tag{C51}$$

where we have separated the bipartite 2d square lattice into even/odd sublattices with vertices $\Lambda_e, \Lambda_o$. At $h \neq 0$, the field breaks the on-site $\mathbb{Z}_2$ symmetry generated by

$$X_{\text{tot}} = \prod_{i \in \Lambda} X_i. \tag{C52}$$

However, defining the shift operators $T_{x,y}$ via their actions on single-site operators:

$$T_x^\dagger A_{(x,y)} T_x := A_{(x+1,y)}, \tag{C53a}$$

$$T_y^\dagger A_{(x,y)} T_y := A_{(x,y+1)}, \tag{C53b}$$

where $(x,y) \in \Lambda = \mathbb{Z}_{L_0} \times \mathbb{Z}_{L_0}$, we notice that we can build a $\mathbb{Z}_{L_0} \times \mathbb{Z}_{L_0}$ symmetry group for $H$ generated by $\ell_x$ and $\ell_y$, defined as

$$\ell_{x,y} := T_{x,y} X_{\text{tot}}. \tag{C54}$$

Indeed it is easy to check that $[H, \ell_{x,y}] = 0$, $[\ell_x, \ell_y] = 0$ and $\ell_x^{L_0} = \ell_y^{L_0} = 1$ (since $L_0$ is even). We also note that $[T_{x,y}, X_{\text{tot}}] = 0$. As a consequence, $H$ no longer has a $\mathbb{Z}_2$ on-site symmetry but it does have a hybrid translation-flavor symmetry group $G = \mathbb{Z}_{L_0} \times \mathbb{Z}_{L_0}$.

Let us now establish that the system above is also stable to $G$-symmetric perturbations that are local in space (note that this is a stronger requirement than the locally-symmetric perturbations used in previous theorems):

**Theorem C.14.** *Suppose q-local $V = \sum_{F \subset \Lambda} V_F$ where $V_F$ is supported on a connected subset $F$ of the 2d square lattice, with local strength $\epsilon$ given by* (B11). *Suppose $V$ is symmetric under $G$ described above:*

$$\ell_x^\dagger V \ell_x = \ell_y^\dagger V \ell_y = V. \tag{C55}$$

*If $\epsilon \leq c_2 J$, the low-energy eigenstates of the perturbed $H = H_0 + V$ with energy* (B34) *have $G \to G/\mathbb{Z}_2$ $(R, c_4 e^{-c_3 L_0}, L_*)$-SSB with respect to the usual bottleneck structure of $H_0$ in Proposition C.4, where $G/\mathbb{Z}_2$ is generated by $(2, 0)$ and $(1, 1)$ translations. Here the constants $c_2, c_3, c_4 > 0$ are determined by $q, R$.*

*Proof.* We aim to prove a condition like global QPC (B32), following the strategy of Theorem C.3 that compares energies of a state $|\psi\rangle$ with another state $\ell_A |\psi\rangle$ obtained from a truncated symmetry action. Without loss of generality, the broken symmetry generator is $(1, 0)$ and so we will be interested in dividing up the space of configurations into cosets separated by $\ell_x$. Define truncated symmetry action $\ell_A$ for any connected $A \subset \Lambda$ as

$$\ell_A = \left(\prod_{i \in A} X_i\right) \cdot \mathcal{T}_A, \tag{C56}$$

where $\mathcal{T}_A$ cyclically permutes $A$ along $x$ direction and acts trivially on the complement $A^c$; see Fig. 3(b) for an illustration. More precisely, if $A = A_1 \sqcup A_2 \sqcup \cdots$ where each $A_y$ is $A$ restricted to one $y$ coordinate, $\mathcal{T}_A = \otimes_y \mathcal{T}_{A_y}$ where $\mathcal{T}_{A_y}$ cyclically permutes $A_y$.

Although $\mathcal{T}_A$ involves long-range permutations, the key point is that for an operator $\mathcal{O}$ in the interior of $A$, $\mathcal{T}_A^\dagger \mathcal{O} \mathcal{T}_A = T_x^\dagger \mathcal{O} T_x$. Then since $V$ is $\ell$-symmetric,

$$\ell_A^\dagger V \ell_A - V = \sum_{F: F \cap \partial A \neq \emptyset} \ell_A^\dagger V_F \ell_A + \sum_{F: F \cap \partial A = \emptyset} \ell^\dagger V_F \ell - \sum_F \ell^\dagger V_F \ell$$

$$= \sum_{F \subset \Lambda: F \cap \partial A \neq \emptyset} (\ell_A^\dagger V_F \ell_A - \ell^\dagger V_F \ell) \tag{C57}$$

only involves terms on the boundary of $A$.[12] Therefore, the bound (C4) on the $V$-energy difference after applying $\ell_A$ also holds here.

On the other hand, (C2) for the new truncated symmetry $\ell_A$ does not hold for all states with $E_B |\mathbf{z}\rangle = n |\mathbf{z}\rangle$ where $n$ is close to $L_B$. The reason is that even for $n = L_B - 1$, the boundary bits $\partial A$ in $\mathbf{z}$ could be half 0 and half 1 that almost perfectly mismatch with their neighbors outside of $A$. Applying $\prod_{i \in A} X_i$ deletes the DWs, but truncated translation $\mathcal{T}_A$ may make the DWs reappear by moving 0s to locations of 1s.

Nevertheless, the previous analysis can be tightened to not requiring (C2) for *all* states $\mathbf{z}$: We only require it for $\mathbf{z}$ that are close to the closed-DW-loop state with $E_B = L_B$, because in the expanded eigenstate equation (B21), $|\psi_2\rangle$ only consists of those bitstrings that can be mapped by local operators to $|\psi_1\rangle$ with $E_B = L_B$. This argument iterates to $|\psi_n\rangle$ for all $n \leq n_*$, while $|\psi_{n_*+1}\rangle$ consists of bitstrings with $E_B = L_B - 1$ that are not close to the closed-DW-loop case.

Formally, consider all $\mathbf{z}$ with

$$\text{Ham}(\mathbf{z}, \mathbf{z}_0) \leq \xi L_B, \tag{C58}$$

for some $\mathbf{z}_0$ with $E_B |\mathbf{z}_0\rangle = L_B |\mathbf{z}_0\rangle$, where $\xi > 0$ is a sufficiently small constant independent of $L_0$. Without loss of generality, suppose $\mathbf{z}_0$ has all ones in boundary $\partial A$ and all zeros in the outer boundary $\partial A^c$, $\mathbf{z}$ is also all ones in $\partial A$ except for at most $\xi L_B$ sites. Since permutation $\mathcal{T}_A$ moves at least half of the bits in $\partial A$ to remain in the boundary $\partial A$, there are $\geq \frac{1}{2}|\partial A| - \xi L_B$ bits in $\partial A$ that are zero in $\ell_A \cdot \mathbf{z}$. Therefore, $\geq \frac{1}{2}|\partial A| - \xi L_B$ checks in loop $B$ become satisfied, because each site in $\partial A$ is acted on by at least one check. On the other hand, $\leq 4\xi L_B$ checks in loop $B$ are

---

[12] Here it is crucial that $F$ is connected, which is why we stressed this point in the statement of the theorem. In particular, $V_F$ must act on $\partial A$ if it acts on sites in $A$ and $A^c$. In contrast, in Theorem C.3, $F$ can be disconnected given that $V$ is locally symmetric. It would be interesting, if it is possible, to generalize our method to allow for a locally-symmetric perturbation even when the symmetries involve translations.

satisfied in $\mathbf{z}$, because each of the $\xi L_B$ bits different from $\mathbf{z}_0$ is touched by at most 4 checks. The energy difference is then

$$H_0(\mathbf{z}) - H_0(\ell_A \cdot \mathbf{z}) \geq \frac{1}{2}|\partial A| - \xi L_B - 4\xi L_B \geq c_\xi J L_B \tag{C59}$$

where constant $c_\xi > 0$ if $\xi$ is sufficiently small. Here we have used $|\partial A| \geq \frac{1}{4} L_B$.

Following the proof of Theorem C.3, (C57) and (C59) establishes the energy estimate (B32) with $\Delta = c_\xi J - 2\epsilon > 0$ for any state $\psi$ reachable in Hamming distance $\xi L_B$ to the case where $B$ is an entire DW-loop, as long as $\epsilon/J$ is sufficiently small. As discussed two paragraphs above, this version of global QPC is sufficient to enable an analogous proof to Corollary B.9, where the relevant states after expanding the eigenstate equation is precisely those reachable states; for brevity we will not rehash every step here. Ultimately, the conclusion of Corollary B.9, and furthermore the conclusion of Theorem C.9 that there is SSB, then follow with a different set of constants $c_2, c_3, c_4$ in the current context. $\square$

## Appendix D: Metastability of false vacua

Recently, [40] developed a systematic theory for many-body metastable quantum states based on the existence of a state $|\psi\rangle$ which is a "local gapped ground state" – i.e. all local operators acting on $|\psi\rangle$ either stabilize it or raise its average energy by at least $\Delta$. One then finds that there is a "prethermal" stability of this metastable state $|\psi\rangle$. The existence of a local gap was crucial for the applicability of the technical methods used in [40]. In this appendix, we will show that the Peierls methods developed above can be used to present a generalized theory of many-body quantum metastability and slow false vacuum decay, even when the metastable state is "locally gapless". Indeed, we will need no assumptions other than a QPC on a sufficiently large (but finite!) length scale in order to derive the existence of a false vacuum with a slow decay rate.

### D.1. Metastability from quantum Peierls condition with finite $L$

Our results will rely on the locality of dynamics:

**Definition D.1.** *Consider a subset $B \subset \Lambda$ and $A = \{i \in \Lambda : \mathsf{d}(i, B) \leq R_B\}$. We say Hamiltonian $H$ satisfies $(t_{\mathrm{LR}}, \delta_{\mathrm{LR}})$-local-simulatability for the sets $A, B$, if the following holds: For any evolution time $|t| \leq t_{\mathrm{LR}}$, any operator $\mathcal{O}_B$ acting on $B$ is evolved approximately by $H_A$ that is sum of terms in $H$ supported inside $A$, with error*

$$\left\| e^{itH} \mathcal{O}_B e^{-itH} - e^{itH_A} \mathcal{O}_B e^{-itH_A} \right\| \leq \delta_{\mathrm{LR}} \|\mathcal{O}_B\|. \tag{D1}$$

(D1) can be viewed as a non-standard form of Lieb-Robinson bound. See e.g. [55] for a review on Lieb-Robinson bounds and a proof of (D1) in strictly local systems where $t_{\mathrm{LR}} = R_B/(2v)$ for some constant $v$ and $\delta_{\mathrm{LR}}$ is exponentially small in $R_B$. The above more general form also applies to long-range interactions [64–67].

**Theorem D.2** (Metastability from local quantum Peierls condition). *Consider Hamiltonian $H$ that satisfies $(t_{\mathrm{LR}}, \delta_{\mathrm{LR}})$-local simulatability for the sets $A, B$ chosen in Definition D.1. Suppose that $H_A$ obeys a $(\Delta, \theta, L, R)$-QPC with respect to a quantum bottleneck structure. For any well $k$, let $\mathcal{P}_k^A$ be a projector (acting on $A$) onto $M$ eigenstates of $P_k H_A P_k$ with energy $E$ obeying (B14). Here the $M$ states can be arbitrarily chosen in the set of all such states. Then for any density matrix $\rho$ such that reduced density matrix $\rho_A$ obeys*

$$\mathrm{Tr}\left(\mathcal{P}_k^A \rho_A\right) > 1 - \delta', \tag{D2}$$

*for any operator $\mathcal{O}_B$ acting in $B$ that is normalized ($\|\mathcal{O}_B\| = 1$),*

$$\left| \mathrm{Tr}\left( \rho \left( e^{iHt} \mathcal{O}_B e^{-iHt} - e^{iP_k H_A P_k t} \mathcal{O}_B e^{-iP_k H_A P_k t} \right) \right) \right| \leq 8|t|\sqrt{M}\delta + 4\sqrt{\delta'} + \delta_{\mathrm{LR}}, \tag{D3}$$

*for all times $|t| \leq t_{\mathrm{LR}}$. Here $\delta$ is given by (B17) with $\|H\|$ replaced by $\|H_A\|$.*

*If furthermore $\mathcal{P}_k^A \mathcal{O}_B \mathcal{P}_k^A \geq c_k > 0$, then*

$$\mathrm{Tr}\left(\rho e^{iHt} \mathcal{O}_B e^{-iHt}\right) \geq c_k(1 - \delta') - 2\sqrt{\delta'} - 8|t|\sqrt{M}\delta - \delta_{\mathrm{LR}} \tag{D4}$$

The key point of Theorem D.2 is that even if we have a finite-sized notion of metastability on a fixed scale $L$, there are still important and observable consequences. Local observables $\mathcal{O}_B$ will evolve as if the dynamics is restricted to within a bottleneck $k$ for timescales set by $R_B$ from Lieb-Robinson bound. In particular, if $\mathcal{O}_B$ has signature in well $k$, that signature is preserved approximately up to time $t_{\text{LR}}$.

Although $R_B$ and $L$ are finite constants, we will show in an explicit example (Theorem D.3) that that $R_B \sim e^L$ is achievable in quantum Ising models. This will establish the non-perturbatively slow decay of gapless false vacua obeying a QPC.

*Proof of Theorem D.2.* The local simulatability (D1) approximates the evolution by $H_A$:

$$\left|\text{Tr}\left(\rho e^{iHt}\mathcal{O}_B e^{-iHt}\right) - \text{Tr}\left(e^{-iH_At}\rho_A e^{iH_At}\mathcal{O}_B\right)\right| = \left|\text{Tr}\left(\rho\left(e^{iHt}\mathcal{O}_B e^{-iHt} - e^{iH_At}\mathcal{O}_B e^{-iH_At}\right)\right)\right| \leq \delta_{\text{LR}}, \quad (D5)$$

where we have evaluated $\text{Tr}(\rho\, e^{iH_At}\mathcal{O}_B e^{-iH_At}) = \text{Tr}(e^{-iH_At}\rho_A e^{iH_At}\mathcal{O}_B)$ by thinking of time-evolving $\rho_A$ rather than $\mathcal{O}_B$. From (D2) and Theorem B.6, we know that $\rho_A$ mostly consists of $\delta$-almost eigenstates of $H_A$ with $\delta$ given by (B17). Let $\rho_A(t) := e^{-iH_At}\rho_A e^{iH_At}$, $\widetilde{\rho}_A(t) := e^{-iH_At}\mathcal{P}_k^A \rho_A \mathcal{P}_k^A e^{iH_At}$, we have

$$\|\rho_A(t) - \widetilde{\rho}_A(t)\|_1 = \|\rho_A - \widetilde{\rho}_A\|_1 = \left\|(1-\mathcal{P}_k^A)\rho_A + \mathcal{P}_k^A\rho_A(1-\mathcal{P}_k^A)\right\|_1$$
$$\leq 2\left\|(1-\mathcal{P}_k^A)\rho_A\right\|_1 = 2\left\|(1-\mathcal{P}_k^A)\sqrt{\rho_A}\cdot\sqrt{\rho_A}\right\|_1 \quad \text{(Triangle inequality and } \|\mathcal{P}_k^A\mathcal{O}\|_1 \leq \|\mathcal{P}_k^A\|\|\mathcal{O}\|_1 = \|\mathcal{O}\|_1)$$
$$\leq 2\left\|(1-\mathcal{P}_k^A)\sqrt{\rho_A}\right\|_2 \|\sqrt{\rho_A}\|_2 \quad \text{(Hölder's inequality)}$$
$$= 2\left[\text{Tr}\left((1-\mathcal{P}_k^A)\rho_A\right)\text{Tr}(\rho_A)\right]^{1/2} \leq 2\sqrt{\delta'} \quad (D6)$$

using (D2). Similarly,

$$\left\|e^{-iP_kH_AP_kt}\rho_A e^{iP_kH_AP_kt} - \rho'_A(t)\right\|_1 \leq 2\sqrt{\delta'} \quad (D7)$$

where

$$\rho'_A(t) := e^{-iP_kH_AP_kt}\mathcal{P}_k^A\rho_A\mathcal{P}_k^A e^{iP_kH_AP_kt} = e^{-i\mathcal{P}_k^AH_A\mathcal{P}_k^At}\mathcal{P}_k^A\rho_A\mathcal{P}_k^A e^{i\mathcal{P}_k^AH_A\mathcal{P}_k^At} \quad (D8)$$

because $\mathcal{P}_k^A$ commutes with $P_kH_AP_k$.

To compare $\widetilde{\rho}_A(t)$ with $\rho'_A(t)$, we apply Duhamel identity $e^{t\mathcal{A}} = e^{t\mathcal{B}} + \int_0^t dt' e^{(t-t')\mathcal{A}}(\mathcal{A}-\mathcal{B})e^{t'\mathcal{B}}$:

$$\widetilde{\rho}_A(t) - \rho'_A(t) = -i\int_0^t dt' e^{-i(t-t')H_A}\left[H_A - \mathcal{P}_k^AH_A\mathcal{P}_k^A, \rho'_A(t')\right]e^{i(t-t')H_A}. \quad (D9)$$

Taking trace norm of (D9) and using triangle inequality, we get

$$\|\widetilde{\rho}_A(t) - \rho'_A(t)\|_1 \leq \int_0^t dt' \left\|\left[H_A - \mathcal{P}_k^AH_A\mathcal{P}_k^A, \rho'_A(t')\right]\right\|_1$$
$$= \int_0^t dt' \left\|\left[(1-\mathcal{P}_k^A)H_A\mathcal{P}_k^A + \mathcal{P}_k^AH_A(1-\mathcal{P}_k^A), \rho'_A(t')\right]\right\|_1 \quad ((1-\mathcal{P}_k^A)\rho'_A(t') = 0)$$
$$\leq 4\int_0^t dt' \left\|(1-\mathcal{P}_k^A)H_A\mathcal{P}_k^A\right\| \|\rho'_A(t')\|_1 \quad \text{(Submultiplicativity)}$$
$$\leq 4\int_0^t dt' \left\|(1-\mathcal{P}_k^A)[H_A, \mathcal{P}_k^A]\right\| \cdot 1 \quad ((1-\mathcal{P}_k^A)\mathcal{P}_k^A = 0)$$
$$\leq 4|t|\left\|[H_A, \mathcal{P}_k^A]\right\|. \quad (D10)$$

Let $(E_m, \psi_m)$ $(m = 1, \cdots, M)$ be the eigenstates of $P_kH_AP_k$ in $\mathcal{P}_k^A$:

$$\left\|[H_A, \mathcal{P}_k^A]\right\| = \left\|\sum_{m=1}^M \left(|r_m\rangle\langle\psi_m| - |\psi_m\rangle\langle r_m|\right)\right\| \quad (|r_m\rangle := (H_A - E_m)|\psi_m\rangle)$$
$$\leq 2\left\|\sum_{m=1}^M |r_m\rangle\langle\psi_m|\right\| \quad \text{(Triangle inequality)}$$
$$\leq 2\left(\sum_{m=1}^M \langle r_m|r_m\rangle\right)^{1/2} \quad (\psi_m \text{ are orthogonal})$$
$$\leq 2\sqrt{M}\delta \quad (D11)$$

from (B17).

Combining (D6), (D7), (D10) and (D11) using triangle inequality,

$$\left\|e^{-iP_k H_A P_k t}\rho_A e^{iP_k H_A P_k t} - \rho_A(t)\right\|_1 \leq 8|t|\sqrt{M}\delta + 4\sqrt{\delta'}, \tag{D12}$$

which leads to (D3) using (D5) and $|\text{Tr}(\mathcal{OO}')| \leq \|\mathcal{O}\|\,\|\mathcal{O}'\|_1$.

To show (D4), we use the fact that

$$\|\rho_A(t) - \rho'_A(t)\|_1 \leq 8|t|\sqrt{M}\delta + 2\sqrt{\delta'} \tag{D13}$$

using (D6), (D10) and (D11). Since $\rho'_A$ is supported in $\mathcal{P}_k^A$ in which $\mathcal{O}_B \geq c_k$,

$$\text{Tr}\left(\mathcal{O}_B \rho'_A(t)\right) \geq c_k \text{Tr}\left(\rho'_A\right) = c_k \text{Tr}\left(\mathcal{P}_k^A \rho\right) \geq c_k(1-\delta'). \tag{D14}$$

Combining (D5), (D13) and (D14) yields (D4). □

We also remark that the result above allows us to apply our results to study local observables in a straightforward fashion, in systems in the strict thermodynamic limit, e.g. random bond Ising models where $\Lambda = \mathbb{Z}^d$. One simply restricts to regions $A$ of arbitrarily large (but finite) size.

### D.2. False vacuum decay in the random bond Ising model

As an important application of the general framework for finite-size quantum bottlenecks discussed above, we now present a new proof that the false vacuum decays non-perturbatively slowly in the quantum Ising model. This new proof will also apply to a gapless false vacuum, e.g. in the random bond Ising model, which is beyond the regime of validity of the formalism developed in [40].

**Theorem D.3.** *Consider the same random bond Ising model $H_0$ as Theorem C.11 with constants $0 \leq J_1 < \Delta' < J_2$. Under any q-local perturbation $V$ of local strength $\epsilon$ in (B11), for any constant $R > 0$, $H = H_0 + V$ satisfies $(\frac{\Delta}{2}, \theta, c_1\frac{\Delta}{\epsilon}, R)$ QPC if $p < p_{\rm c}, \epsilon < \epsilon_{\rm c}$ and*

$$L_0 \leq \exp\left(c_2 \frac{\Delta}{\epsilon}\right). \tag{D15}$$

*Here $\Delta, \theta, p_{\rm c}, \epsilon_{\rm c}, c_1, c_2 > 0$ are constants determined by $J_1, J_2, \Delta', q, R$.*

*Furthermore, suppose $V$ is 2d-local in the sense that each $F$ in (B11) is connected in the 2d lattice. Even if (D15) does not hold, for any ball region $B$ of radius*

$$c_3 \frac{\Delta}{\epsilon} \leq r \leq \exp\left(c_2 \frac{\Delta}{2\epsilon}\right) \tag{D16}$$

*there exists a local observable $\mathcal{O}_B$ with $\|\mathcal{O}_B\| = 1$ supported in $B$ such that there are two initial states $\rho_\pm$ with*

$$\pm\text{Tr}\left(\rho_\pm e^{iHt}\mathcal{O}_B e^{-iHt}\right) \geq \frac{1}{2}, \qquad \forall |t| \leq \exp\left(c_4 \frac{\Delta}{\epsilon}\right). \tag{D17}$$

*Here $c_3, c_4$ are constants determined by $J_1, J_2, \Delta', q, R$.*

*Proof.* We use the same bottleneck structure as Proposition A.7 with one difference: The largest DW-loop size in the two wells $W_1, W_2$ is $L = c_1/\epsilon$ that depends on perturbation strength rather than the total system size. The number of bottleneck indicators is still $|\mathcal{B}| \leq e^{\theta L}$ using (A19) and (D15), where constant $\theta$ is determined by $c_1, c_2, R$.

From the proof of Theorem C.11, we know that as long as $p$ is sufficiently small ($< p_{\rm c}$ determined by $\theta, J_1, J_2, \Delta'$), $H_0$ satisfies (C2) for all bottleneck indicators with some constant $\Delta > 0$. Following the proof of Theorem C.3, it remains to show

$$\langle\psi|\,\ell_A^\dagger V \ell_A - V\,|\psi\rangle \leq L_B \frac{\Delta}{2} \tag{D18}$$

in order to prove QPC for $H$ with a slightly smaller $\frac{\Delta}{2}$ parameter. (D18) holds because

$$\langle\psi|\ell_A^\dagger V \ell_A - V|\psi\rangle \leq \left\|\ell_A^\dagger V \ell_A - V\right\|$$

$$= \left\|\sum_{F:F\cap A\neq\emptyset} \ell_A^\dagger V_F \ell_A - V_F\right\| \quad (\ell_A \text{ acts inside } A)$$

$$\leq \sum_{i\in A}\sum_{F\ni i} \left\|\ell_A^\dagger V_F \ell_A - V_F\right\| \quad \text{(Triangle inequality)}$$

$$\leq \sum_{i\in A}\sum_{F\ni i} 2\|V_F\| \leq 2|A|\epsilon$$

$$\leq 2c_3 L_B^2 \epsilon \leq L_B \frac{\Delta}{2}. \tag{D19}$$

Here in the last line, we have used the fact that a $L_B$-size loop encircles an area bounded by $|A| \leq c_3 L_B^2$ for some constant $c_3$, and that $L_B \epsilon \leq 4RL\epsilon = 4Rc_1$ is chosen to be a sufficiently small constant. We have thus proven $(\frac{\Delta}{2}, \theta, c_1 \frac{\Delta}{\epsilon}, R)$ QPC for $H$.

We then apply Theorem D.2 to show the second part of the claim.

First, there exists an observable $\mathcal{O}_B$ with $\|\mathcal{O}_B\| = 1$ such that

$$P_1 \mathcal{O}_B P_1 \geq \frac{2}{3} > -\frac{2}{3} \geq P_2 \mathcal{O}_B P_2. \tag{D20}$$

The reason is that with a sufficiently large constant $c_3$, $r \gg L$ so that we can tell which region is the connected sea of 0s solely from region $B$ for any bitstring in $P_1$: If a DW is only partially contained in $B$, it separates $B$ into regions and we know there is a single largest region which is its outside. One can therefore define a observable $\mathcal{O}_B$ diagonal in the $z$-basis to tell the bit value in the connected sea, which returns $\pm 1$ for the two wells and satisfies (D20).

We then choose region $A$ to be a $L_0' \times L_0'$ box enclosing $B$ with

$$R_B = \frac{1}{2}\exp\left(c_2 \frac{\Delta}{\epsilon}\right) - r - 1 \tag{D21}$$

so that $L_0'$ satisfies (D15). Note that $R_B > 0$ for sufficiently small $\epsilon_c$ due to the upper bound of $r$ (D16). Furthermore, the QPC condition implies $\delta$-almost eigenstates with

$$\delta \leq c_5 (L_0')^2 \exp\left[-c_6 L \log \frac{\epsilon_c'}{\epsilon}\right]$$

$$\leq c_5 \exp\left(2c_2 \frac{\Delta}{\epsilon}\right) \exp\left[-c_6 c_1 \frac{\Delta}{\epsilon} \log \frac{\epsilon_c'}{\epsilon}\right]$$

$$\leq \exp\left(-c_4' \frac{\Delta}{\epsilon}\right) \tag{D22}$$

from (B17), where $c_4', c_5, c_6, \epsilon_c' > 0$ are constants and we have chosen $\epsilon_c = \frac{1}{2}\epsilon_c'$ to be sufficiently small.

Since $V$ is 2d-local, $H$ satisfies a Lieb-Robinson bound with a finite velocity and exponential-decaying tails, and therefore $(c_{\text{LR}} R_B, 0.1)$-local simulatability for some constant $c_{\text{LR}}$ [55].

We then choose $\rho_\pm$ to be the lowest eigenstate of $P_1 H_A P_1$ ($P_2 H_A P_2$), tensored with arbitrary state outside of $A$. Note that the corresponding energies $E$ obey (B14) because it is just $E_{k,\min}$. (D4) then implies (D17) with a sufficiently small $\epsilon_c$ using $c_k = \frac{2}{3}, \delta' = 0, M = 1, \delta_{\text{LR}} = 0.1$, (D22) and a sufficiently small constant $c_4 > 0$ so that $|t| \leq c_{\text{LR}} R_B$ and $|t|\delta$ is exponentially small in $\Delta/\epsilon$. $\square$

Theorem D.3 will generalize in a straightforward way to $d$ dimensions with a $\exp\left[c\left(\frac{\Delta}{\epsilon}\right)^{d-1}\right]$ lifetime for the false vacuum, although we have not gone into details of the Peierls conditions for higher-dimensional models in this paper. It would be interesting if our methods could be modified to prove a tight $\exp\left[c\left(\frac{\Delta}{\epsilon}\right)^d\right]$ lifetime that matches false vacuum decay in non-random Ising models [40].

---


}

\clearpage
\begingroup

\pdfpagewidth=612bp
\pdfpageheight=792bp

\pdfhorigin=0pt
\pdfvorigin=0pt

\newcount\supppage
\supppage=1

\loop
  \pdfximage page \supppage {supp.pdf}%
  \shipout\vbox to 792bp{%
    \offinterlineskip
    \hbox to 612bp{%
      \pdfrefximage\pdflastximage
      \hss
    }%
    \vss
  }%
  \advance\supppage by 1
\ifnum\supppage<29
\repeat

\endgroup

\end{document}